\documentclass[twocolumn, prd, aps,superscriptaddress,preprintnumbers,tightenlines,showpacs,nofootinbib,eqsecnum,amsfonts,amsmath]{revtex4}

\usepackage{epsfig}
\usepackage{graphics}
\usepackage{graphicx}
\usepackage{amsmath,amssymb,mathrsfs}
\usepackage{amsfonts}
\usepackage{xcolor}
\usepackage{wasysym}
\usepackage{times}
\usepackage{mathptmx}
\usepackage{gensymb}
\usepackage{appendix}
\usepackage{mathrsfs}

\begin{document}

\newcommand{\be}{\begin{equation}}
\newcommand{\ee}{\end{equation}}
\newcommand{\ber}{\begin{eqnarray}}
\newcommand{\eer}{\end{eqnarray}}
\newcommand{\bea}{\begin{eqnarray}}
\newcommand{\eea}{\end{eqnarray}}
\newcommand{\ie}{i.e.}
\newcommand{\dt}{{\rm d}t}
\newcommand{\df}{{\rm d}f}
\newcommand{\dtheta}{{\rm d}\theta}
\newcommand{\dphi}{{\rm d}\phi}
\newcommand{\rhat}{\hat{r}}
\newcommand{\iotahat}{\hat{\iota}}
\newcommand{\phihat}{\hat{\phi}}
\newcommand{\hc}{{\sf h}}
\newcommand{\etal}{\textit{et al.}}
\newcommand{\balpha}{{\bm \alpha}}
\newcommand{\bbeta}{{\bm \psi}}
\newcommand{\fmerg}{f_{1}}
\newcommand{\fring}{f_{2}}
\newcommand{\fcut}{f_{3}}
\newcommand{\rmi}{{\rm i}}
\def\rd{{\textrm{\mbox{\tiny{RD}}}}}
\def\qnr{{\textrm{\mbox{\tiny{QNR}}}}}
\newcommand{\A}{\mathcal{A}}
\newcommand{\NS}{\mathrm{NS}}
\newcommand{\CC}{\mathcal{C}}

\newcommand{\sascha}[1]{\textcolor{violet}{\textit{Sascha: #1}}}
\newcommand{\mdh}[1]{\textcolor{red}{\textit{Mark: #1}}}
\newcommand{\patricia}[1]{\textcolor{blue}{\textit{Patricia: #1}}}

\newcommand{\UIB}{Departament de F\'isica, Universitat de les Illes Balears, 
Crta. Valldemossa km 7.5, E-07122 Palma, Spain}
\newcommand{\Cardiff}{School of Physics and Astronomy, Cardiff University, Queens Building, CF24 3AA, Cardiff, United Kingdom}


\title{Towards models of gravitational waveforms from generic binaries: A simple approximate 
mapping between precessing and non-precessing inspiral signals}

\author{Patricia Schmidt}
\affiliation{\Cardiff}

\author{Mark Hannam}
\affiliation{\Cardiff}

\author{Sascha Husa}
\affiliation{\UIB}


\begin{abstract}
One of the greatest theoretical challenges in the build-up to the era of second-generation 
gravitational-wave detectors is the modeling of generic binary waveforms. We introduce an 
approximation that has the potential to significantly simplify this problem. We show that generic 
precessing-binary inspiral waveforms (covering a seven-dimensional space of 
intrinsic parameters) can be 
mapped to a two-dimensional space of non-precessing binaries, characterized by the mass 
ratio and a single effective total spin. The mapping consists of a time-dependent rotation of
the waveforms into the quadrupole-aligned frame, and 
is extremely accurate (matches $> 0.99$ with parameter biases in the total spin of 
$\Delta \chi \leq 0.04$), even in the case of transitional precession. 
In addition, we demonstrate a simple method to construct hybrid 
post-Newtonian--numerical-relativity precessing-binary waveforms in the
quadrupole-aligned frame, and provide evidence that our approximate mapping can be used
all the way to the merger. Finally, based on these results, we outline a general proposal for the 
construction of generic waveform models, which will be the focus of future work.  
\end{abstract}


\maketitle

\section{Introduction}
\label{sec:intro}

The second-generation laser interferometric gravitational-wave detector
Advanced LIGO is 
planned for first commissioning in 2014, and to reach design sensitivity in subsequent
years~\cite{Abbott:2007kv,Shoemaker:aLIGO,2010CQGra..27h4006H}; 
Advanced Virgo~\cite{Accadia:2011zz,aVIRGO} and 
Kagra~\cite{Somiya:2011me} are expected to follow soon 
after. Current estimates
of astrophysical event rates predict that the first direct detection of gravitational waves will occur in 
that time frame~\cite{Abadie:2010cf}. 
The coalescence of two black holes is among the strongest known gravitational-wave sources and a likely 
candidate for one of the first detections. 
The detection and subsequent analysis of gravitational waves relies strongly on the accuracy 
and completeness of theoretical waveform models. 
For black-hole binaries, this includes the inspiral, merger and ringdown of the final black hole, and 
current models combine information from analytic approximation methods and 
numerical-relativity (NR) simulations~\cite{Ohme:2011rm}.

To date, a number of theoretical inspiral-merger-ringdown (IMR) waveform models exist for 
non-spinning binaries 
and configurations where the spin angular momentum is either aligned or anti-aligned with the orbital angular 
momentum (a summary of these models is given in Ref.~\cite{Ohme:2011rm}).
But most astrophysical binary systems are expected to have arbitrary spin configurations, 
which lead to complicated
precession effects. Although there does exist one preliminary precessing-binary IMR 
model~\cite{Sturani:2010yv}, the modeling 
of generic binaries remains a serious challenge.

The complicated structure of precessing-binary waveforms suggests that in order to construct
accurate IMR waveform models, we may need to produce numerical
simulations that densely sample a seven-dimensional parameter space. At first glance, this 
does not seem feasible on the timescale of second-generation GW detectors (i.e., within the 
next ten years), although valiant efforts are underway~\cite{NRAR_web}. 

In this paper we introduce an approximation that has the potential to dramatically simplify the 
modeling of precessing-binary waveforms. Motivated by the results of our previous 
work~\cite{Schmidt:2010it}, we show that the seven-dimensional space of 
intrinsic physical parameters of generic precessing-binary waveforms can be mapped to a 
\emph{two-dimensional} space
of non-precessing waveforms, parametrized by the mass ratio and an effective total spin 
parameter. The mapping consists of transforming the precessing-binary waveforms into a 
``co-precessing'' frame of reference, described by three rotation angles $\{\gamma(t), \beta(t),
\epsilon(t)\}$. This is the ``quadrupole-aligned'' (QA) frame that we introduced 
in~\cite{Schmidt:2010it}.
The waveform modeling problem is then factorized into two much smaller problems:
(1) the construction of a non-precessing-binary model (and candidates for such a model 
already exist \cite{Ajith:2009bn,Santamaria:2010yb,Pan:2011gk}), and 
(2) the construction of a model for the rotation angle functions 
$\{\gamma(t), \beta(t), \epsilon(t)\}$ with respect to the binary's seven physical parameters, which 
we expect can itself be further simplified. In this paper we do not address the (still very large)
task of producing a model for the rotation angle functions, and we leave the behavior of the signal during
the ringdown for future work. Here we restrict ourselves to an outline of 
the approximate mapping between precessing-binary and non-precessing-binary waveforms
and test its validity on a series of inspiral waveforms generated by post-Newtonian (PN) theory. 

Generic binary systems undergoing quasi-circular inspiral are characterized by nine intrinsic 
physical parameters: the binary's total mass $M=m_1+m_2$, the mass ratio 
$q=m_2/m_1$  (we adopt the convention that $m_1< m_2$), and the six spin components 
$\vec{S}_1$ and $\vec{S}_2$. The total mass of the system 
sets the overall scale and can be factored out. The individual masses $m_1$ and $m_2$ 
are uniquely determined given $M$ and $q$. 

In the most simple cases, the two black holes are either not spinning, or their spin angular momenta 
are (anti)-aligned with the orbital angular momentum $\vec{L}$. In these cases 
the inspiral motion is confined to a fixed, time-independent plane. 
The orbital frequency of the motion then grows monotonically with time and, when the GW signal 
is decomposed into spin-weighted spherical harmonics,
most of the emitted gravitational energy is contained in the dominant harmonics, the 
$(l=2,|m|=2)$-modes. In the most general configurations, however, the two black-hole 
spins are not aligned with the orbital angular momentum vector. Now the inspiral motion is 
no longer confined to a fixed plane. 
In these cases precession occurs, specifically two types: precession of the instantaneous 
orbital plane as well as precession of the spin vectors. The now far more complex motion 
is reflected in the emitted radiation in the form of strong amplitude modulations, which depend on 
the relative orientation of the binary towards the observer, and as a contribution to the 
binary's phase evolution. These effects are illustrated further in Sec.~\ref{sec:precession}.

We have previously shown that precessing-binary waveforms take a far simpler form when
transformed into the quadrupole aligned (QA) frame~\cite{Schmidt:2010it}.
In a nutshell, the QA frame approximately follows the instantaneous orbital plane of the binary. 
In this frame the binary is 
essentially viewed ``face-on'' throughout the course of its evolution. 
Note that this frame corresponds to a co-rotating, accelerated frame of reference. In this 
``co-precessing'' frame the amplitudes of the waveform modes as well as their frequency 
evolution are significantly simplified and most of the energy 
is emitted in the $(l=2,|m|=2)$ modes, just as in a non-precessing binary. 
In fact, in this accelerated frame the mode structure of a non-precessing binary appears to be restored 
(see Fig. 12 in Ref.~\cite{Schmidt:2010it} for an NR example). It was this observation that suggested the
idea that we pursue in this paper, that QA- and non-precessing-binary waveforms may agree well in
both amplitude \emph{and} phase. Note that a related frame, defined by the direction of the Newtonian
orbital angular momentum, was introduced in Ref.~\cite{Buonanno:2002fy}, along with the observation
that the precession-induced phase oscillations can be removed in this ``precessing frame''. The
key new result in this paper, beyond the use of the QA frame (which can be determined from the 
GW signal alone), is the simple mapping between precessing-frame waveforms and non-precessing-binary
waveforms, as we describe below.

In the context of gravitational-wave searches and parameter estimation, waveforms from different binary 
configurations are most strongly characterized by their phase evolution, i.e., their rate of inspiral. When the
black holes are widely separated their motion can be described well by PN methods, and in PN theory we
see that the leading-order influence of the spin on the inspiral rate and the phase evolution is 
the spin-orbit coupling, which is due to a sum of the components of the black-hole spins 
parallel to the orbital angular momentum~\cite{Kidder:1995zr}. 
If the binary precesses, then the precession will 
introduce both secular 
and oscillatory changes in the phase, but in the QA frame, where the precession has been removed to some
extent, we expect to recover the underlying orbital phase evolution, which will be similar
to that for a non-precessing binary. Since the leading-order spin effects on the phase arise from the 
total black-hole spin, it is possible to make an approximate parametrization of non-precessing binaries
by a single effective total spin parameter, $\chi_{\rm eff}$, and this idea has been used in 
both inspiral~\cite{Ajith:2011ec} and IMR~\cite{Santamaria:2010yb} models. In this work we focus on 
inspiral PN models, and so we will use the same effective spin parameter as in Ref.~\cite{Ajith:2011ec};
see Sec.~\ref{sec:pncomp}.
For complete IMR waveforms
other parameterizations have been found to work better \cite{Ajith:2009bn,Santamaria:2010yb}, but in this work 
we restrict ourselves to PN inspiral waveforms.

Our hypothesis, then, is that precessing-binary waveforms can be approximately mapped to 
non-precessing-binary waveforms, and that the equivalent non-precessing-binary signal is 
parameterized by the mass ratio and $\chi_{\rm eff}$.
It is the goal of this paper to quantify the accuracy of that approximation. 
Our approach is to consider a selection of PN inspiral
precessing-binary waveforms, and to match them against a family of non-precessing-binary signals, to
determine the best-match value of $\chi_{\rm eff}$. We can then see how well these values agree with 
our expectation, and the level of agreement with the best-match waveform.
We use PN waveforms because they allow us to study the long inspiral regime, and they 
are far more computationally convenient to produce than numerical simulations of only the last
$\sim 10$ orbits before merger.  

In Sec.~\ref{sec:precession} we will describe some of the general features of precessing-binary 
systems before giving a brief summary of the PN inspiral waveforms in Sec.~\ref{sec:pn}. 
We provide a technical summary of the 
quadrupole-alignment procedure in Sec.~\ref{sec:qa}. We perform our study of PN waveforms in 
Sec.~\ref{sec:pncomp} and 
demonstrate the efficacy of our approximate mapping not only for a large number of cases that exhibit 
simple precession, but also for an example configuration that undergoes transitional precession. 
In Sec.~\ref{sec:hybrids} we show that the QA frame greatly simplifies the construction of hybrid PN-NR waveforms for
precessing configurations, and discuss the potential extension of our approach to full IMR waveforms. 
Finally, we sketch a procedure to construct generic IMR models and discuss the issues that must first
be overcome. 

\section{Precessing-binary waveforms}
\label{sec:wfs}
\subsection{General features}
\label{sec:precession}

In this section we summarize the main features of precessing-binary systems and illustrate the 
effects of precession on the gravitational-wave signal. For a comprehensive discussion of 
precessing-binary systems we refer the reader to Refs.~\cite{Apostolatos:1994mx,Kidder:1995zr},
which remain the standard references in the field.

In non-spinning or spin-aligned cases the normal to the orbital plane, i.e., the Newtonian 
orbital angular momentum $\hat{L}_N$, is well-defined and does not evolve and provides a
useful direction with which to characterize the dynamics. 
In the presence of precession any such characteristic direction becomes 
time-dependent. But one nearly-fixed direction in the binary configuration does exist: the direction 
of the total angular momentum remains close to its limit when the binary has infinite separation. 
We denote this the ``asymptotic total angular momentum direction'' 
$\hat{J}_{-\infty} \equiv \hat{J}(t\rightarrow -\infty)$.
(We will use a hat to denote unit vectors). 
This is analogous to standard Newtonian solid-body mechanics, where the system rotates about the axis 
defined by the total angular momentum, which is a natural fixed direction. This is still true for Newtonian 
and first-order post-Newtonian binary systems. When spin effects are included, starting at 1.5PN order, and
in full General Relativity, this natural direction of rotation still exists, but it is no longer fixed. 
The direction of the total angular momentum is now time-dependent. It evolves, but in cases with
small precession, and for large separations, it describes a 
precession cone that is rather small compared to all of the other time-dependent  directions, 
like the orbital angular momentum or the spin directions~\cite{Brown:2012gs}.
In a few special cases, where the orbital and spin angular momenta are nearly equal and 
opposite and the total angular momentum passes through zero during the inspiral, the direction 
of the total angular momentum changes rapidly; this is called transitional precession.

The complex dynamics in precessing systems is reflected in the gravitational-wave signal.
The precession introduces amplitude modulations but also contributes to the phasing. 
Furthermore, the power  contained in various gravitational-wave spherical harmonic modes, 
defined with respect to a fixed coordinate system, is significantly affected by 
precession, as power is transferred to the modes that were subdominant in the non-precessing 
configurations. 
Nonetheless, since the GW signal is to first approximation produced by the acceleration of the two bodies 
in orbit, the bulk of the energy is emitted along the direction of the orbital angular momentum $\hat{L}$. 
This is the idea behind quadrupole-alignment: if we track the direction of the maximum energy emission, 
then we will also be tracking the orbital precession. 

This also provides insight into the signal observed from a fixed direction. The orbital angular 
momentum $\hat{L}$ precesses around $\hat{J}_{-\infty}$, and on average we expect the 
bulk power to be radiated in the direction of 
$\hat{J}_{-\infty}$. This point is discussed in more detail in~\cite{O'Shaughnessy:2012vm}. 
If a generic precessing binary now happens to be 
ideally oriented towards some static observer, i.e., the line-of-sight and 
$\hat{J}_{-\infty}$ coincide, only small 
amplitude modulations will be observed in the gravitational waveform, since the relative orientation 
between the observer and the least-precessing axis of the binary does not change much. On the other 
hand, if the observer's orientation does not coincide with this most stable axis, then they will 
observe strong modulations.

\begin{figure}
\begin{center}
\includegraphics[width=80mm]{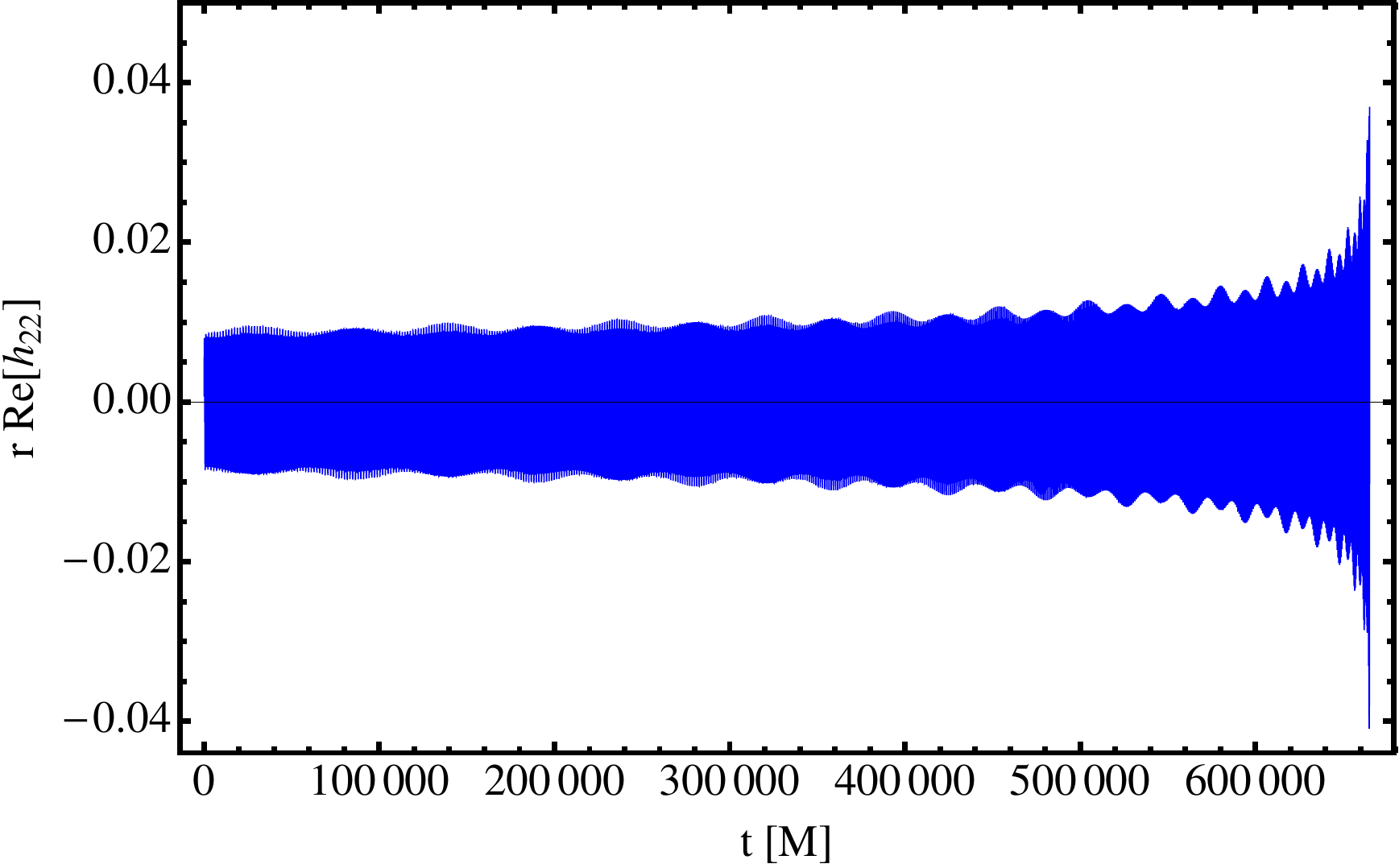}
\includegraphics[width=80mm]{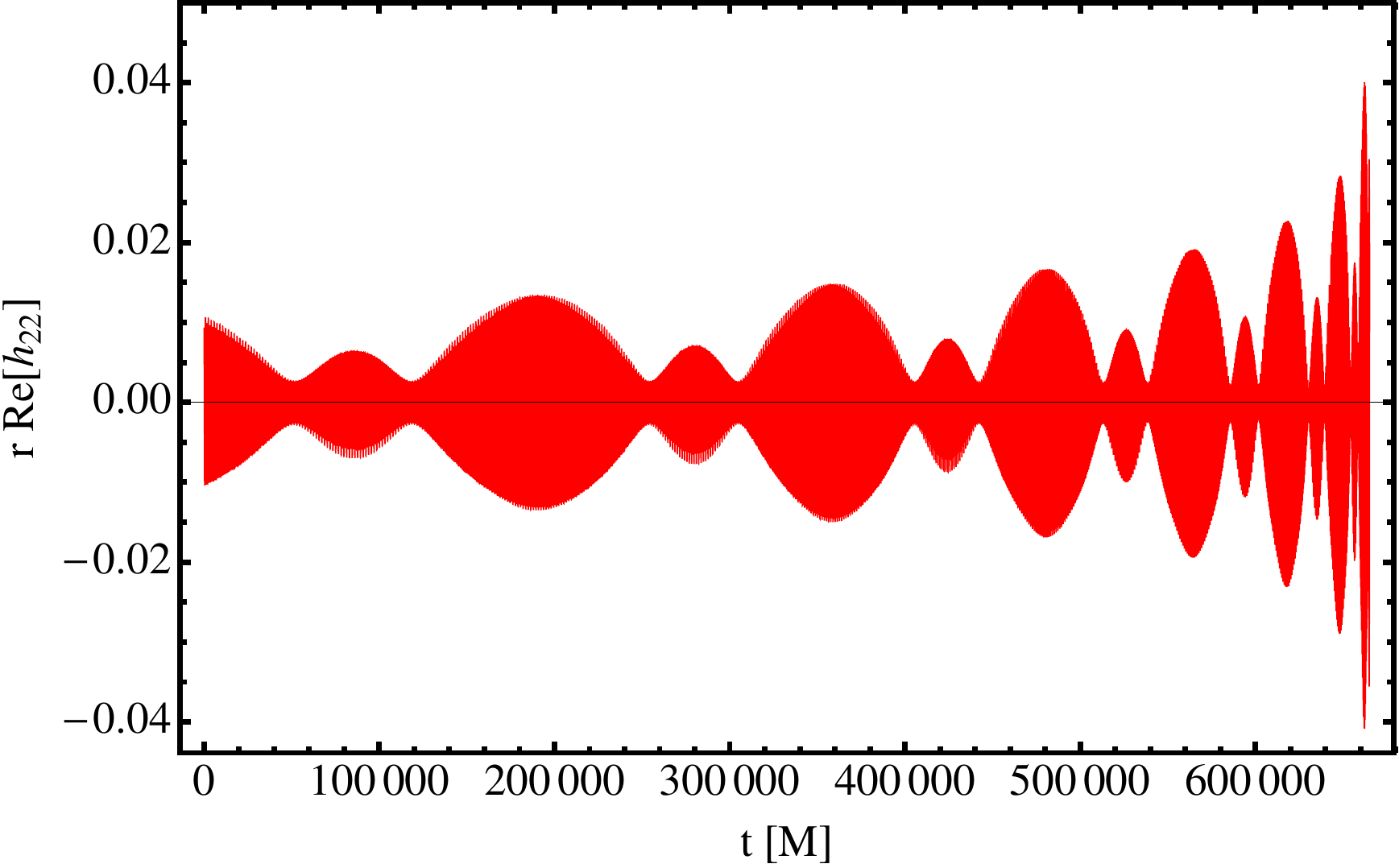}
\caption{The first panel shows the real part of the $(\ell=2,m=2)$ mode with $\hat{J}$ initially 
aligned with $\hat{z}$, for the 1:10 binary described in the text. The second panel shows the 
same quantity, but now with $\hat{L}$ 
initial aligned with $\hat{z}$. It is clear that if we are searching for signals with a monotonically 
increasing amplitude (as with non-precessing binaries), we may easily class the signal in the
lower panel as a glitch in the data.}
\label{fig:real}
\end{center}
\end{figure}

To illustrate this point, we consider a single-spin binary system with mass ratio $1:10$, 
where the larger black hole has an initial spin of $\vec{\chi}_2=(0.75,0,0)$ and the smaller 
one is non-spinning. We expect to see only very few oscillations from directions close to 
$\hat{J}_{-\infty}$. This is illustrated in Fig.~\ref{fig:real}: the first panel shows the real part of the 
post-Newtonian $(2,2)$-mode of the GW strain $h$ (see Sec.~\ref{sec:pn} for details) of the precessing 
binary as seen along the direction of the binary's initial total angular momentum. 
The second panel shows the real part of 
the $(2,2)$-mode for the same configuration, but for an observer whose line-of-sight does not coincide
with the direction of the initial total angular momentum. In this case, the line-of-sight 
coincides with the direction of the Newtonian angular momentum at the beginning of the 
waveform. 
This direction varies with time and crosses the observer's line-of-sight 
after each precession cycle. The amplitude peaks are observed when the maximal emission 
direction points towards the observer. (Note that since the line-of-sight coincides with $\vec{L}$
at $t=0$, the amplitude has a maximum at this time.)

The detectability of a signal in a matched-filter GW search can be estimated based on its 
best match against all of the theoretical signals in the template bank; the match is unity if
the template bank contains precisely that signal, and is zero if the signal is orthogonal to all
of the template-bank waveforms. 
(Match calculations will be described in more detail in Sec.~\ref{sec:pncomp}.)
For the two waveforms shown in Fig.~\ref{fig:real}, the first has
a match against a non-precessing-binary template bank of over 0.97 (which means that 
more than 90\% of such signals would be found in a search), while the second has a best
match below 0.7, meaning that it would most likely either be missed, or classed as a glitch
in the data. These illustrations provide a useful perspective on the results in~\cite{Ajith:2011ec},
where it was shown that spin-aligned GW searches are very likely to miss a large number 
of generic signals at high mass ratios. 
We see that these signals are lost not simply because precessing-binary signals are 
very different from non-precessing-binary signals. Rather, for some orientations they look 
very similar, but the fraction of those orientations
decreases as the precession effects increase.

These examples demonstrate the dramatic difference in the precessing-binary waveforms 
with respect to relative orientation. But we have seen that even in the
$\hat{J}_{-\infty}$ direction modulations remain in the waveform, 
and we will see in Sec.~\ref{sec:qa} that these are further reduced when we go to 
the QA-frame.

\subsection{Post-Newtonian Waveforms}
\label{sec:pn}

In order to produce the precessing post-Newtonian inspiral waveforms used in this analysis, 
we evolved the full PN equations of motion, which were integrated using a Mathematica 
package. In many studies of precessing binaries and in GW search work, it is common to 
use adiabatic inspiral models, for example the ``SpinTaylor'' 
equations, which are the precessing-binary extension of TaylorT4.
But, in order to capture as much of the full physics as possible, we prefer to 
use instead an evolution of the full PN equations of motion, formulated as
the Hamiltonian equations of motion in the standard
Taylor-expanded form, as we have done previously~\cite{Husa:2007rh,Hannam:2010ec,Purrer:2012wy}.
More specifically, we use the non-spinning 3PN accurate Hamiltonian~\cite{Jaranowski:1997ky,Damour:2001bu,Damour:2000kk} (see also~\cite{Blanchet:2000ub,deAndrade:2000gf,Blanchet:2002mb}) and the 3.5PN
accurate radiation flux~\cite{Blanchet:1997jj,Blanchet:2001aw,Blanchet:2004ek}. We add both leading-order~\cite{Barker1970,Barker1974,Barker1979,Kidder:1995zr,Damour:2001tu,Poisson:1997ha}
and next-to-leading order~\cite{Blanchet:2006gy,Faye:2006gx,Damour:2007nc}
contributions to the spin-orbit and spin-spin Hamiltonians, and the
spin-induced radiation flux terms as described in~\cite{Buonanno:2005xu}
(see also~\cite{Kidder:1995zr,Poisson:1997ha}). In addition we include the flux contribution 
due to the energy flowing into
the black holes, which appears at the relative 2.5PN order, as derived in
Ref.~\cite{Alvi:2001mx}. 

The precessing PN waveforms were then generated making use of the explicit formulae 
for the waveform modes $h_{\ell m}$ as given by Eqn. (B1) and (B2) in~\cite{Arun:2008kb}. 
The expression for the $(2,0)$-mode was provided by G. Faye and the $(2,-m)$-modes where
constructed according to Eq. (4.15) in~\cite{Arun:2008kb}.
The positions, momenta and spins of the masses were read off the full PN solution and used 
to generate the parameters for the construction of the precessing waveform modes $h_{2m}$. 
These modes contain only the leading-order spin contributions but higher-order corrections 
are contained in the dynamics, since the Hamiltonian is known to higher order (see above). 
Therefore, even if the $h_{\ell m}$ expressions were evaluated only at quadrupole-order, 
the waveforms would still show many features of
precession, since the dominant contribution to the waveforms is from the motion itself. 
Note that the dynamical calculations are performed in the ADMTT gauge, while the 
mode expressions are written in the harmonic gauge. This inconsistency will introduce
errors into the waveforms, but we do not expect these to be larger than those due to the 
neglect of higher-order PN contributions. 

We have set up the source coordinate system as in Ref.~\cite{Arun:2008kb}, where $\hat{J}_0= (0,0,1)$ 
and defines the total angular momentum direction at the initial separation. 
To achieve this, the PN initial data $\{\vec{q},\vec{p}, \vec{S}_1, \vec{S}_2\}$ were rotated  
by applying a standard rotation matrix about the $y$- and $z$-axes in the Cartesian source frame. 
This is purely a convention as all of the physics is invariant with respect to rotations.  
The system was evolved for $15$M to reduce eccentricity (as done previously in numerical
applications~\cite{Husa:2007rh,Hannam:2010ec,Purrer:2012wy}), and then from an initial 
separation of $D_i=40$M down to a final separation of $D_f=6$M. 

The orbital frequency of the motion is given by the general expression
\begin{equation}
\label{eq:omega}
\vec{\omega}_{orb}=\frac{\vec{q} \times \dot{\vec{q}}}{\|\vec{q}\|^2},
\end{equation}
where $\vec{q}$ is the relative separation of the point masses and $\dot{\vec{q}}$ the relative 
velocity. The Newtonian orbital angular momentum is given by 
\begin{equation}
\label{eq:LN}
\vec{L}_N=\mu(\vec{q} \times \dot{\vec{q}}),
\end{equation}
where $\mu$ denotes the reduced mass 
\begin{equation}
\label{ }
\mu=\frac{m_1 \cdot m_2}{m_1+m_2}.
\end{equation}
The general PN orbital angular momentum vector $\vec{L}$ is given by
\begin
{equation}
\label{eq:L}
\vec{L}=\vec{q} \times \vec{p}.
\end{equation}
Note that $\vec{L}_N$ and $\vec{L}$ differ significantly in the case of  precession since 
$\dot{\vec{q}}$ and $\vec{p}$ are no longer strictly parallel to each other, as explained 
in~\cite{Kidder:1995zr,Schmidt:2010it} unless the two masses $m_1$ and $m_2$ are 
 far apart. In the case 
of precession, the directions of $\vec{L}, \vec{L}_N, \vec{S}_i$ and $\vec{J}$ are all 
time-dependent. The left panel of Fig.~\ref{fig:JLmotion} shows the time evolution of 
$\hat{J}$ and $\hat{L}$.  We see that $\hat{J}$ evolves on 
a much smaller precession cone than the orbital angular momentum. 

The Newtonian orbital angular momentum $\vec{L}_N$ is defined by its polar coordinates 
$\{\iota(t),\alpha(t)\}$, which are measured with respect to the $z$-axis of our Cartesian 
source system. The evolution of these two angles describes the dynamics of the 
instantaneous orbital plane. They are defined by
\begin{eqnarray}
\iota (t) & = & \arccos{\left(\frac{L_{Nz}}{\|\vec{L}_N\|}\right)}, \\
\alpha (t) & = & \arctan{\left(\frac{L_{Ny}}{L_{Nx}}\right)}. 
\end{eqnarray}
The total phase of the binary is then constructed from the following integral~\cite{Arun:2008kb}:
\begin{equation}
\label{eq:phase}
\Phi(t)=\int_0^t (\omega_{orb}(t')-\dot{\alpha}(t')\cos\iota(t'))dt'.
\end{equation}
The physical interpretation of the integral is as follows:
the phase seen by an observer on the $z$-axis (which is the axis that defines our 
mode decomposition of the GW signal) is a combination of the actual motion of
the orbital plane and its projection onto the $xy$-plane of the source frame. This can be
better understood if we first consider a simplified example where 
$\iota$ is constant (the orbital plane is tilted by a fixed angle with respect to $\hat{z}$), 
and $\dot{\alpha}$ is also constant ($\vec{L}_N$ precesses around $\vec{J}$ with a
constant frequency). Then we see that the average observed frequency of the objects' motion
will be larger or smaller than $\omega_{orb}$ depending on the sign of $\dot{\alpha}$.
It is the phase from this observed frequency that $\Phi$ describes.

The symmetric and anti-symmetric spin combinations are constructed directly from the 
solution data:
\begin{eqnarray}
\vec{\chi}_s & = & \frac{1}{2}(\vec{\chi}_1+\vec{\chi}_2), \\
\vec{\chi}_a& = &  \frac{1}{2}(\vec{\chi}_1-\vec{\chi}_2), 
\end{eqnarray}
where the dimensionless spins $\vec{\chi}_i$ are defined from the spin angular momenta $\vec{S}_i$ of each
black hole by 
\begin{equation}
\label{ }
\vec{\chi}_i=\frac{\vec{S}_i}{m_i^2}.
\end{equation}
Once all time-dependent dynamical parameters are constructed, the waveform modes, 
$h_{lm}$, are evaluated. These are most commonly derived by expanding the complex polarization $h$,
\begin{equation}
\label{ }
h=h_+ - ih_\times,
\end{equation}
in the  basis of spin-weighted spherical harmonics ${}^{s}Y_{lm}$ with spin-weight $s=-2$ 
due to the nature of the gravitational field:
\begin{equation}
\label{ }
h_{lm}=\int h(\theta,\varphi) {}^{-2}Y^*_{lm}(\theta,\varphi) d\Omega,
\end{equation}
where $^*$ denotes complex conjugation. For a non-precessing binary this means that if the 
source frame was chosen such that $\hat{L}_N$ is parallel to $\hat{z}$, the quadrupole 
contributions are $h_{22}$ and $h_{2,-2}$. For precessing binaries, $\hat{L}_N$ is not
in general parallel to $\hat{z}$, and hence modes with 
$m \neq |2|$ appear \emph{even at quadrupole order}. They only vanish when $\iota=0$ and 
$\alpha=\pi$. 

\begin{figure}[t]
\begin{center}
\includegraphics[width=80mm]{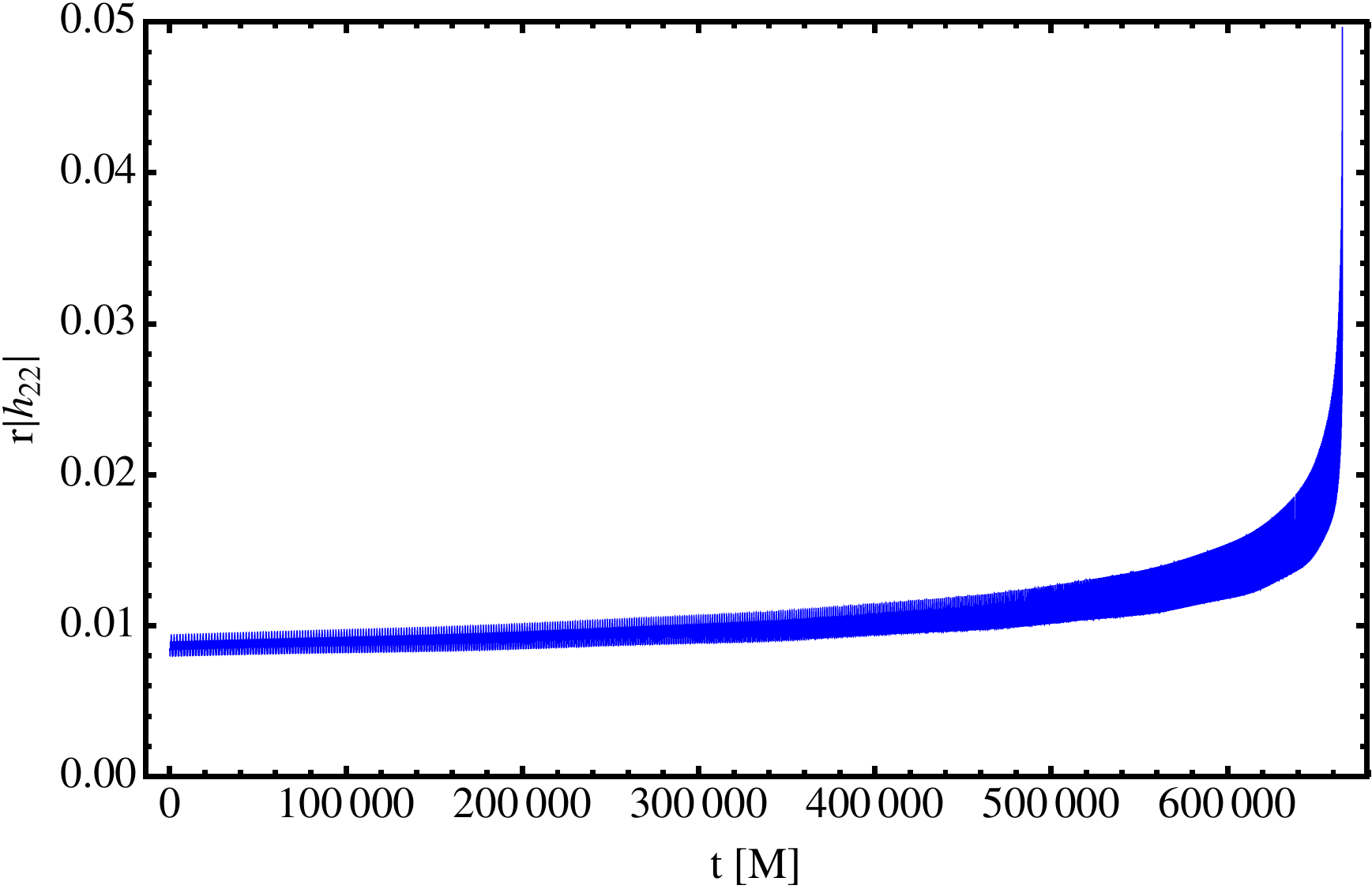}
\includegraphics[width=80mm]{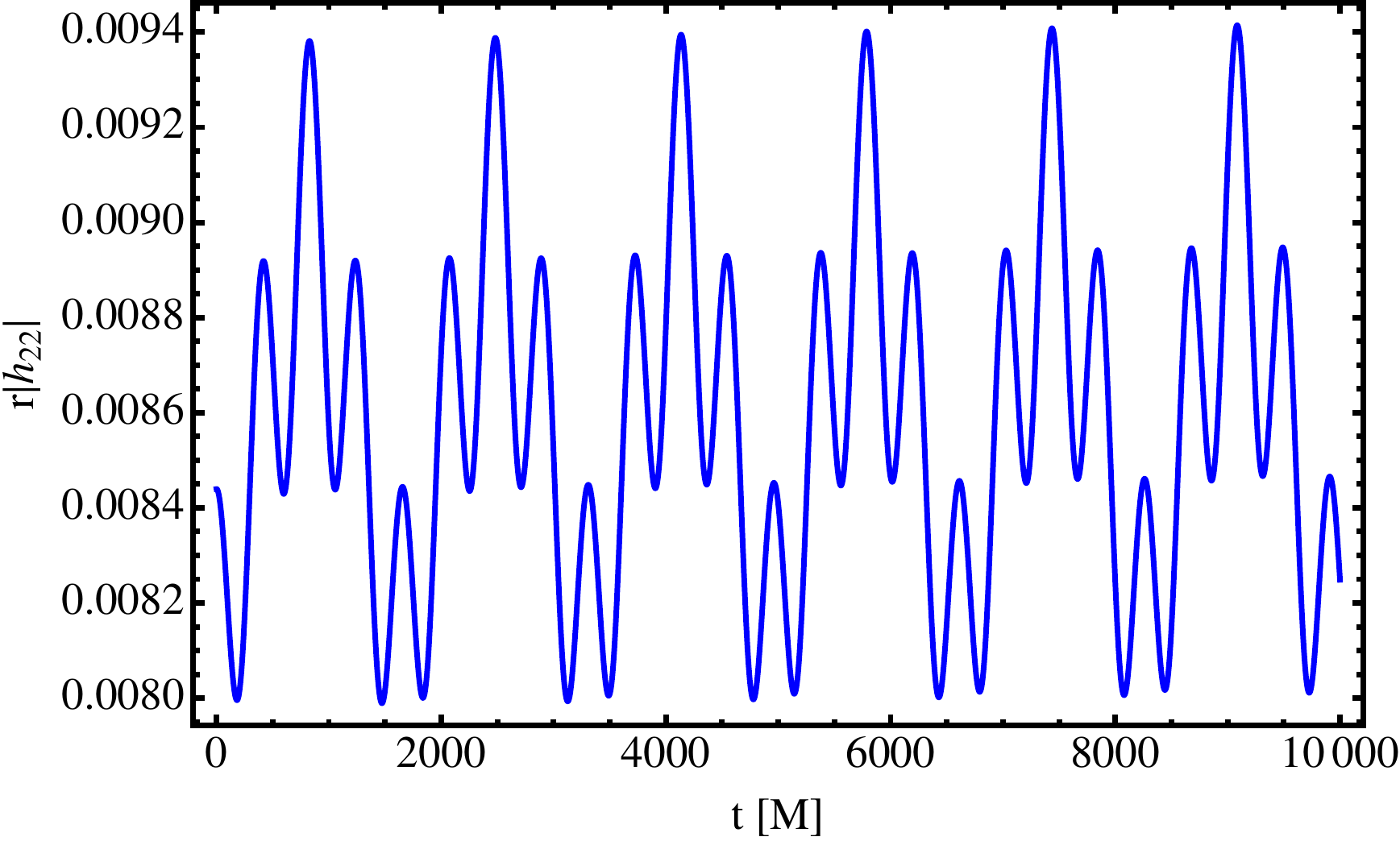}
\caption{The top panel shows the magnitude of the $(2,2)$-mode for the same strongly 
precessing case as in Fig.~\ref{fig:real} over the whole length of the evolution, and over 
a length of the first $10000M$ in the lower panel. The source frame was chosen such that $\hat{J}_0 \simeq (0,0,1)$, and $\vec{L}$ and $\vec{\chi}_2$ 
are initially orthogonal to each other with $\|\vec{\chi}_2\|=0.75$; the smaller black hole is not 
spinning. The close-up of the waveform magnitude over a shorter timescale reveals 
strong amplitude modulations.}
\label{fig:h22J0}
\end{center}
\end{figure}

Schematically, the $h_{lm}$ modes can be written as
\begin{equation}
\label{ }
h_{lm}(t)=f(M,r,\mu,\omega_{orb},\iota,\alpha,\Phi,\vec{\chi}_s,\vec{\chi}_a).
\end{equation} 
The expressions are evaluated for a constant luminosity distance $r$, which is scaled out 
of our results. Fig.~\ref{fig:h22J0} shows the magnitude of the $(2,2)$-mode for the same 
precessing case as described in Sec.~\ref{fig:real}. Despite this being a strongly precessing case 
$(\vec{S}\cdot \vec{L}=0)$, long-timescale modulations are hardly noticeable. This is because 
a preferred frame was already chosen for the evolution, as described previously. 
Only an observer whose line-of-sight coincides with $\hat{J}_0$ will see a signal of this form. 
The appearance and strength of amplitude modulations strongly depends on the relative 
viewing angle, as illustrated in Sec.~\ref{sec:precession}.

\subsection{Quadrupole-Alignment}
\label{sec:qa}
The idea of quadrupole-alignment is to track the direction of the dominant radiation 
emission. This means that, to leading order, it follows the precessing motion of the 
instantaneous orbital plane. This allows us to significantly simplify the gravitational-wave 
signature by artificially removing the precession of the instantaneous orbital plane and 
describing the signal in a co-rotating way. In a previous study we have found evidence 
that the quadrupole-aligned direction actually tracks the full PN angular momentum direction,
which differs slightly from the normal to the orbital plane.  We will discuss this further in an upcoming 
publication~\cite{Schmidt:2012}.

We introduced the idea of the QA frame in~\cite{Schmidt:2010it} and illustrated its properties 
with reference to numerical-relativity waveforms.
We also specified an explicit algorithm to determine the two time-dependent rotation angles 
$\{\beta(t),\gamma(t)\}$ that specify the direction that maximizes the amplitude of the 
$(\ell=2, |m|=2)$-modes. A third angle, $\epsilon(t)$, which adjusts the phase, was ignored in
our original prescription, but its importance was pointed out in~\cite{Boyle:2011gg}, particularly
in whenever $\beta$ is close to zero and $\gamma$ changes rapidly. 
An alternative algorithm to calculate these angles 
was later given in~\cite{OShaughnessy:2011fx}. 

The alignment itself is based on the general transformation behavior of spin-weighted 
spherical harmonics under coordinate transformations. This allows us to find the instantaneous, 
average direction of maximal emission by transforming the $(\ell=2, |m|=2)$-modes
and averaging over the dominant harmonics. This direction is uniquely defined by two 
angles, $\beta$ and $\gamma$, which are determined by the maximization algorithm 
presented in \cite{Schmidt:2010it}:
\begin{equation}
\label{eq:max}
(\beta_{MAX},\gamma_{MAX})=\max_{\beta,\gamma} \sqrt{\|\tilde{h}_{22}(\beta,\gamma)\|^2+\|\tilde{h}_{2,-2}(\beta,\gamma)\|^2},
\end{equation}
where $\tilde{h}_{22}$ and $\tilde{h}_{2,-2}$ are given by
\begin{eqnarray}
\tilde{h}_{22}(\beta,\gamma)&=&\sum_{m'=-2}^{2}e^{-im'\gamma(t)} d^2_{m'2}(-\beta(t))h_{2m'}(t), \\ 
\tilde{h}_{2,-2}(\beta,\gamma)&=&\sum_{m'=-2}^{2}e^{-im'\gamma(t)} d^2_{m',-2}(-\beta(t))h_{2m'}(t), \nonumber \\
\end{eqnarray}
where $d^2_{m'm}$ denote the Wigner d-matrices \cite{Wigner1959,Goldberg:1967}. 
The maximization determines the two 
Euler angles $\beta_{MAX}$ and $\gamma_{MAX}$.
In general, the transformation of spin-weighted spherical harmonics involves three degrees of 
freedom and, as noted in~\cite{Boyle:2011gg}, the third angle can be provided by the analog of 
Eq.~(\ref{eq:phase}), given the other two angles:
\begin{equation}
\label{eq:thirdangle}
\epsilon(t)=-\int \dot{\gamma}_{MAX}(t')\cdot \cos\beta_{MAX}(t') dt'.
\end{equation} We may set $\epsilon(0) = 0$ without loss of generality.

Once all three time-dependent angles $(\beta_{MAX}, \gamma_{MAX}, \epsilon)$ 
have been determined, the dominant quadrupole-aligned mode can then be written as
\begin{equation}
\label{eq:h22QA}
h_{22}^{QA}(t)=e^{-2i\epsilon(t)}\sum_{m'=-2}^{2}e^{-im'\gamma_{MAX}(t)} d^2_{m'2}(-\beta_{MAX}(t))h_{2m'}(t).
\end{equation}
All other QA modes can be constructed as well, as long as the $h_{lm}$-modes for a 
given $l$ are known. One may see that this transformation differs slightly from the one 
presented in~\cite{Schmidt:2010it}. This is because the numerical-relativity waveforms
presented there are related to the PN waveforms in this work by an overall complex 
conjugation.

The three angles $\{\gamma,\beta,\epsilon\}$ define a standard Euler rotation of the reference
frame: a rotation by $\gamma$ about the $z$-axis, followed by a rotation by $\beta$ about the 
$y$-axis, followed by another rotation by $\epsilon$ about the (new) $z$-axis. This is 
important to bear in mind if we consider the reverse procedure to ``wrap up'' a 
QA waveform back into its original precessing-binary form. In that case, the reverse
procedure consists of applying the rotations in the opposite order, i.e., the same 
procedure but with $\{\gamma,\beta,\epsilon\} \rightarrow \{-\epsilon,-\beta,-\gamma\}$.

Although we expect QA waveforms to be useful tools in standardizing the representation 
of precessing waveforms for comparison purposes (as in, for example, 
Ref.~\cite{Hannam:2009hh} for equal-mass nonspinning waveforms), and in waveform modeling, 
they do not correspond to a signal seen by a gravitational-wave detector. 
The QA waveforms are the waves as seen in a very specific accelerated ``co-precessing'' frame. 
One of the consequences of this frame choice is  that the usual relationship 
$\Psi_{4} = -\ddot{h}$ no longer holds, as can be seen by inspection of Eq.~(\ref{eq:h22QA}). 
Hence, in order to obtain quadrupole-aligned Weyl scalar modes, one has to construct 
the precessing modes first and then transform them into the quadrupole-aligned counterparts.
Note also that the QA angles will differ slightly when calculated from either $h$ or 
$\Psi_4$ (this point is also made in~\cite{Ochsner:2012dj}; the $\Psi_4$ angles tend to be
smoother than the $h$ angles).

\begin{figure}
\begin{center}
\includegraphics[width=80mm]{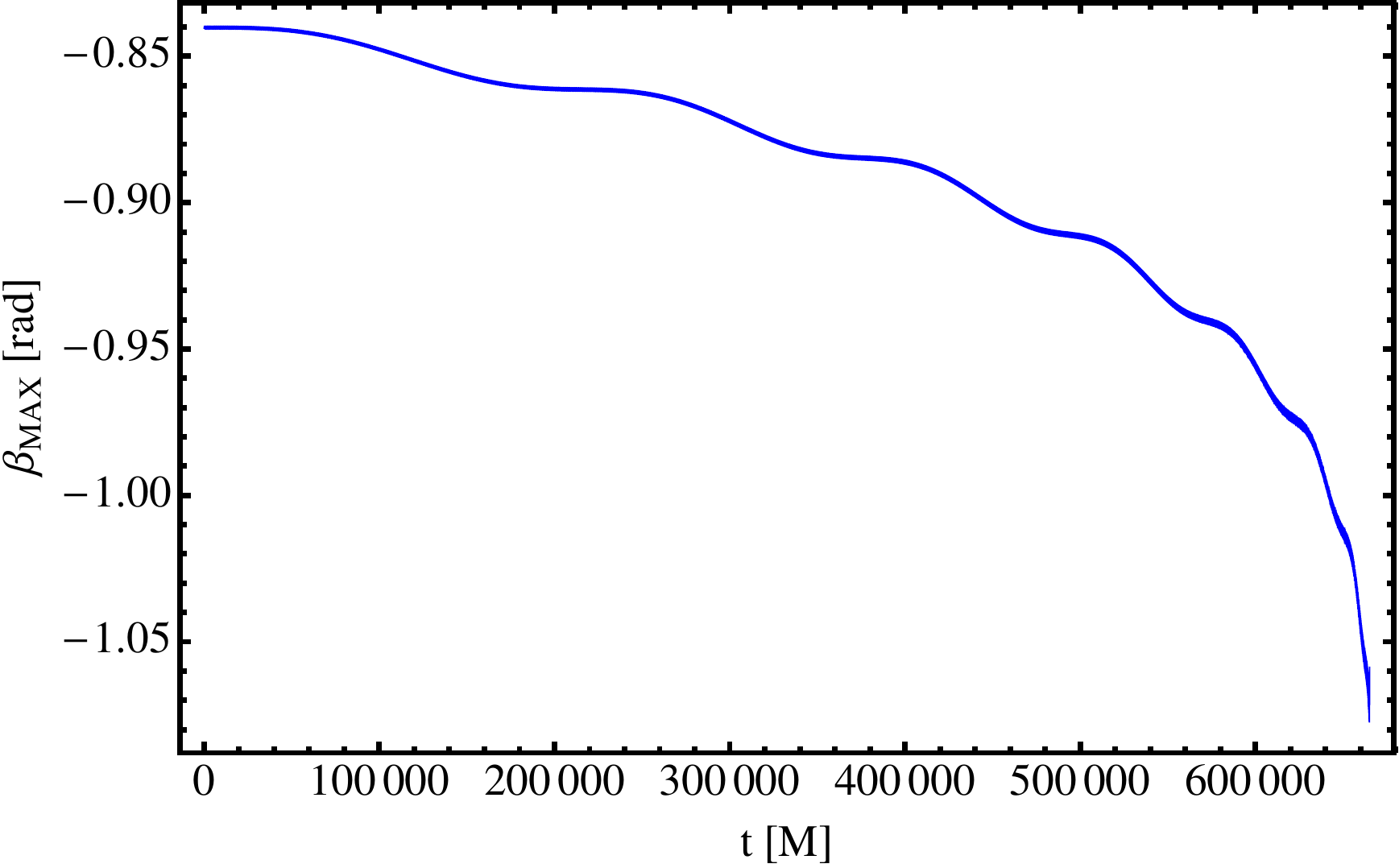}
\includegraphics[width=80mm]{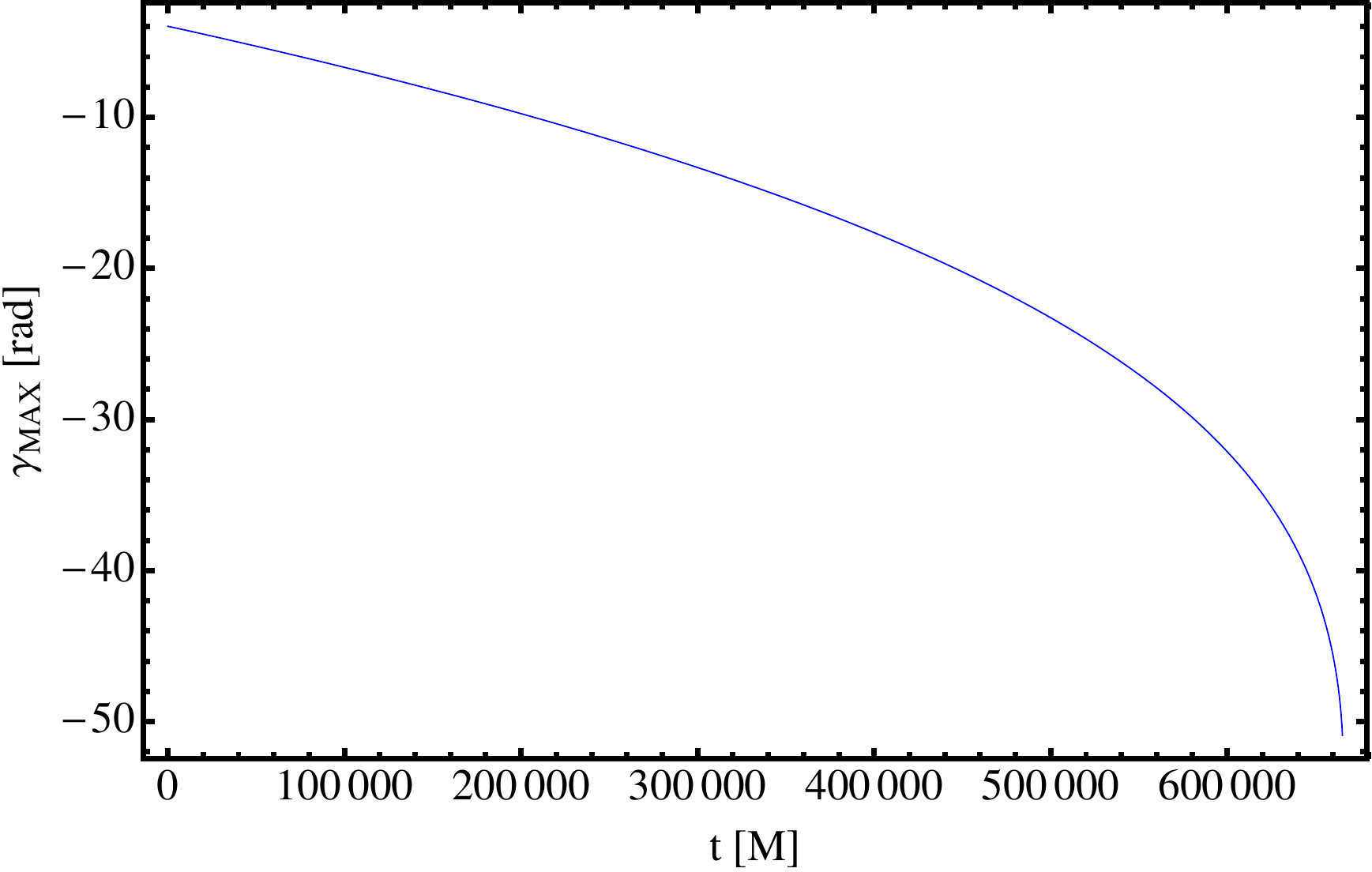}
\caption{The two panels show the angles found in an example of the maximisation routine. 
The first panel shows the inclination angle $\beta_{MAX}$ vs. time,
the second panel shows the azimuth $\gamma_{MAX}$ vs. time over 
the full length of the PN inspiral.}
\label{fig:QAangles}
\end{center}
\end{figure}

To leading PN order, the recovered angles correspond to the inverse Newtonian 
angles $(\iota, \alpha)$, but higher order contributions in the wave amplitudes lead to a deviation 
from those angles, which is consistent with the results from the pure numerical analysis 
in~\cite{Schmidt:2010it}. From that work we expect the angles we find to correspond to 
the smooth evolution of $\hat{L}$ in the limit of a complete description. The angles found by the 
maximization routine are shown in Fig.~\ref{fig:QAangles}. They deviate slightly from the
inverse Newtonian ones $(-\iota, -\alpha)$ 
due to higher-order PN contributions to the mode amplitudes but this difference is not visible
over the scale of the plots.
If we were to use only the quadrupole contribution of the 
$h_{\ell m}$ expressions, then we would recover the direction of $\hat{L}_N$. 

\begin{figure}
\begin{center}
\includegraphics[width=80mm]{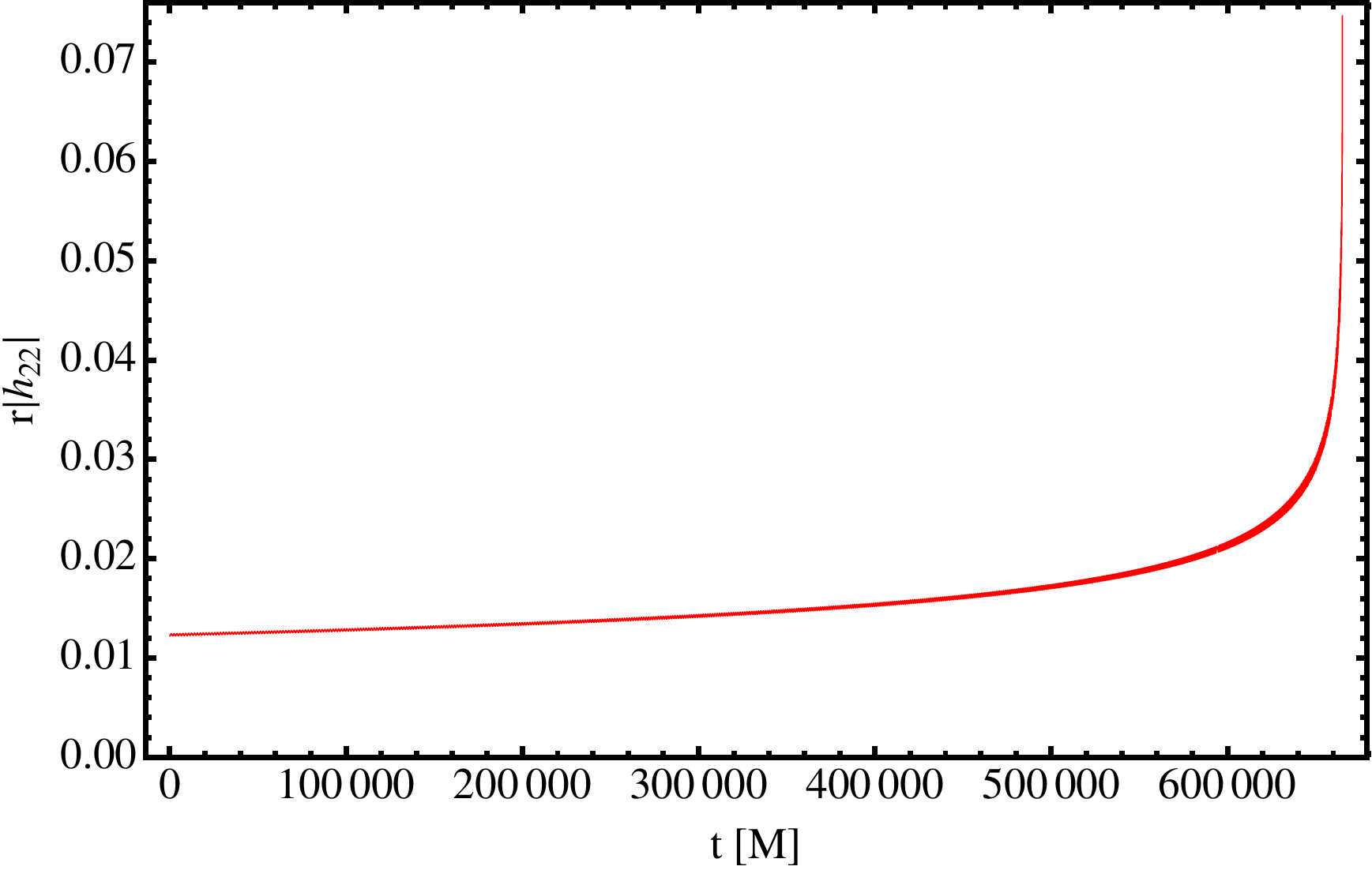}
\includegraphics[width=80mm]{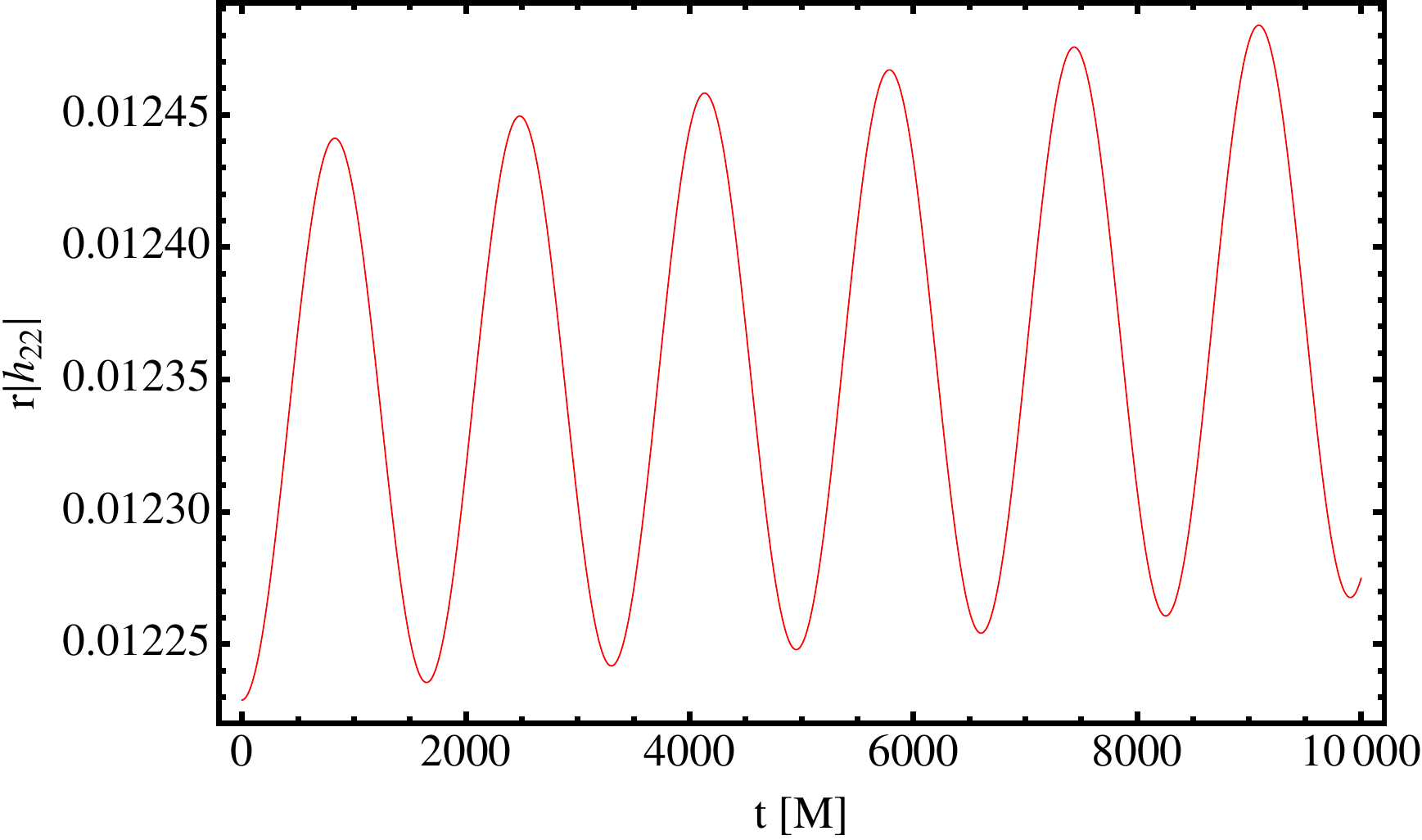}
\caption{QA magnitude for the $q=10$ configuration considered in Figs.~\ref{fig:real} and
\ref{fig:h22J0}. The top panel shows the complete waveform, while the lower panel 
zooms in on the first $10000\,M$. We see that the oscillations in the amplitude have been
reduced and simplified from those in Fig.~\ref{fig:h22J0}.}
\label{fig:QAamp}
\end{center}
\end{figure}

Once the three Euler angles are determined, those are then used to reconstruct the 
QA modes. Fig.~\ref{fig:QAamp} shows the quadrupole-aligned $(2,2)$-mode for the 
configuration shown in Figs.~\ref{fig:real} and \ref{fig:h22J0}. 

In the next section we will present a detailed study of how these simplified QA waveforms 
compare with corresponding non-precessing cases.

\section{Results}
\label{sec:pncomp}

The aim of this section is to test and quantify the accuracy of our hypothesis that 
generic inspiral signals can be mapped 
onto non-precessing counterparts (see Sec.~\ref{sec:intro}). 
Numerical-relativity waveforms are too short for a real inspiral comparison and, moreover, 
it is computationally very expensive to produce a large number of accurate numerical precessing 
waveforms. Instead, we have restricted this analysis to PN waveforms to allow a more detailed 
study for a large subset of the precessing parameter space.

First, we will take a look at simple precession and consider a range of spin configurations for
two mass ratios. The first is $q=3$ and includes the configuration of 
the numerical case that we studied in~\cite{Schmidt:2010it}. The second is $q=10$,
motivated by the observation that precession effects become more significant for
higher mass ratios; see, for example, Eqn.~(2.11) in~\cite{Kidder:1995zr}, and the results
presented in~\cite{Ajith:2011ec}. 
We will show that the mapping works extremely well; the non-precessing waveforms that agree
best with each QA-transformed precessional configuration follow closely the $\chi_{\rm eff}$ 
parameter that we discussed in Sec.~\ref{sec:intro} (and will elucidate further below) 
and agree with them with matches above 0.99. 
Finally, as the most challenging test of our hypothesis, we look at a case of transitional precession. 

This study covers only a small range of the full precessing-binary parameter space, but the 
configurations were carefully chosen to test the hypothesis for varying spin magnitudes and for
two mass ratios within the range that is likely to be treated in IMR models in the near future,
i.e., cases which can also be realized in current numerical simulations to high accuracy. 

From the PN expressions for the phase evolution of the binary~\cite{Kidder:1995zr}, 
we see that the 
dominant spin contribution is proportional to the projection of each spin vector onto the orbital angular momentum, 
$(\vec{S}_i\cdot \vec{L})$. We characterize the degree of spin-alignment with $\kappa_i$, which is defined
as the angle between $\vec{S}_i$ and $\vec{L}$,
\begin{equation}
\label{eq:kappa}
\kappa_i=\arccos \left( \frac{\vec{S}_i\cdot \vec{L}}{\|\vec{S}_i\|\|\vec{L}\|}\right).
\end{equation}
When the spin interaction is restricted to the leading order spin-orbit coupling and radiation 
reaction is switched off,  each $\kappa_i$ is conserved and is a constant of the motion~\cite{Kidder:1995zr}. 
When radiation reaction is included and, to a lesser degree, when higher order spin interactions 
are included, $\kappa_i$ has been observed to show only small variation in time.

The agreement or disagreement between two waveforms is mainly due to their phasing. 
If the inspiral rate is significantly 
different, two waveforms are not expected to agree very well. For the QA waveforms, 
the precession of the orbital plane has been factored out, but the physical spins are, of course, 
present and contribute to the phase evolution. Thus, in general, we expect the best 
comparison waveform to be from a spinning-black-hole binary. 
At leading PN spin-order, where only the leading order spin-orbit terms contribute,
each spin contribution is proportional to                                      
$\cos \kappa_i$, and thus
by looking at the leading-order terms, 
we expect that all waveforms with $\cos \kappa_i=0$ map onto non-spinning counterparts, while 
all waveforms with $\cos \kappa_i \neq 0$ map onto spinning waveforms, which can be parameterized 
by an effective total-spin parameter. 
This 2-part leading-order spin term can be represented by a \emph{single} reduced spin parameter~\cite{Ajith:2011ec}:
\begin{equation}
\label{eqn:chieff}
\chi_{\rm eff}=\chi_{sz}+\frac{(m_1-m_2)}{m}\chi_{az}-\frac{76\eta}{113}\chi_{sz},
\end{equation}
where $\eta$ is the symmetric mass ratio. Note that this parameter is not the same effective
spin parameter as introduced in Ref.~\cite{Damour:2001tu}. 
In this work the effective total spin used \emph{is} the reduced spin parameter as defined by Eq.~(\ref{eqn:chieff}).

In our study the non-precessing-binary 
comparison modes were parameterized by $\vec{\chi}_1=\vec{\chi}_2=(0,0,\chi)$.
For each of these cases we have  
$\chi_{\rm eff} = \chi (1 - 76 \eta/113)$.

The first set of configurations was chosen such that $\kappa_i =0$ for the spinning hole, yielding an effective spin of zero. 
The second set was chosen such that all configurations have the same theoretical effective 
spin of $\chi_{\rm eff} =0.5$, but with varying $\kappa_1=\kappa_2$. The details are listed in 
Tab.~ \ref{tab:pn1} and Tab.~\ref{tab:pn2}. 
The PN comparison family with (anti-)aligned spins was generated by the same method as 
the precessing ones, solving the full PN equations of motion and using the same $h_{\ell m}$ 
expressions~\cite{Arun:2008kb}, where $\alpha=\pi$ and $\iota=0$. This ensures that 
the results are not contaminated by differences due to the choice of the PN approximant.

The agreement between two waveforms can be quantified by a single number, the \textit{match} 
$\mathscr{M}$, which corresponds to a noise-weighted inner product (overlap) between 
them~\cite{Cutler94}. 
Since QA waveforms are not in an inertial (detector) frame, and we are interested in quantifying 
the difference between two waveforms independently of a detector, we primarily use the 
white-noise spectrum $S_n(f)=1$.
Match calculations are performed in the frequency domain and hence the FFTs of the time-domain 
waveform modes have to be computed first. The best match between two frequency-domain 
waveforms $h_1(f)$ and $h_2(f)$ is defined as their normalized inner product maximized over 
time and phase shifts 
($\Delta t$ and $\Delta\phi$):
\begin{equation}
\label{ }
\mathscr{M}=\max_{\Delta t,\Delta\phi} \frac{\langle h_1|h_2\rangle}{\sqrt{\langle h_1|h_1\rangle \langle h_2 | h_2\rangle}},
\end{equation} where the inner product is defined by 
\begin{equation}
 \label{eq:scalar_prod}
 \langle h_1|h_2\rangle := 4 \, {\rm Re} \left[ \int_{f_{\rm
         min}}^{f_{\rm max}} 
     \frac{ \tilde h_1(f) 
     \tilde h_2^\ast (f)}{S_n(f)} \, df \right] \, .
 \end{equation} 
 In our examples the PN waveforms are defined in the frequency range 
 $fM \in [0.0018,0.01]$. The upper frequency corresponds to $M \omega \approx 0.06$, which is 
 typical of the frequency at which we would start using NR results in full IMR hybrids; in this
 study we are not interested in the performance of the PN waveforms beyond that frequency.
 Since the matches are calculated with a flat noise spectrum, they 
 are independent of the binary's mass. 
 
Although the QA waveforms are not in a detector's frame of reference, it is also instructive
to calculate matches with respect to realistic detector noise curves. In this case different choices
of binary mass correspond to giving extra weight to different frequency ranges in the 
waveforms, and provide a more stringent test on the robustness of our results. 
We repeated the match calculation for every configuration with
the early Advanced LIGO~\cite{G1000176} and the zero-detuned high-power~\cite{T0900288} 
noise curves. The matches were calculated for masses between $20\,M_{\odot}$ and $50\,M_{\odot}$ in the frequency range between 20\,Hz and 8\,kHz. 

The idea of the comparison is to find the non-precessing waveform as a function of $\chi$ that gives the 
best match with each QA waveform of our study. If our hypothesis holds, then the best-match spin 
$\chi_{\rm BM}$ will be close to the effective spin $\chi_{\rm eff}$. We present our results 
in the following subsections.

\subsection{Simple Precession}
\label{sec:simple}
The first two sets of PN configurations are cases of simple precession. 
For most arbitrary binary configurations, simple precession will occur 
and only a small set of configurations will undergo ``transitional precession'', as it requires fine-tuned physical 
parameters (see \cite{Apostolatos:1994mx} and Sec.~\ref{sec:transitional} below). 
In the case of simple precession, 
the total spin angular momentum $\vec{S}$ precesses around the orbital angular momentum 
vector $\vec{L}$ and both of these vectors precess around the centre of the rather small 
precession cone described by $\vec{J}_0$. This is illustrated in the left panel of 
Fig. \ref{fig:JLmotion}.

Each precessing time-domain waveform was generated with respect to a source frame 
where $\hat{J}_0$ is initially parallel to the $z$-axis. The quadrupole-alignment algorithm 
was then applied to determine the time series of the two Euler rotation angles 
$\{\beta_{MAX}(t),\gamma_{MAX}(t)\}$. Given those, the third angle, $\epsilon(t)$, 
was determined and Eq.~(\ref{eq:thirdangle}) applied to reconstruct the time-domain quadrupole-aligned 
$(2,2)$-mode. \footnote{Higher modes can be reconstructed as well but here we consider 
only the dominant harmonic in the match calculations.}

The first set of configurations tests the mapping hypothesis for a vanishing proposed 
theoretical effective spin $\chi_{\rm eff}=0$, for various spin configurations for the 
two mass ratios $1:3$ and $1:10$. The results in Tab.~\ref{tab:pn1} suggest that the hypothesis 
works very well for single-spin systems with only the smaller black hole spinning. 
In these cases, we obtain best matches $\geqslant 0.99$ for the theoretical $\chi_{\rm eff}$-value 
for both mass ratios. In the reversed cases, i.e., now the larger black hole is spinning, 
the maximal matches are still $\geqslant 0.99$ but we see a small parameter bias of 
$\Delta \chi = 0.02$. If both black holes are spinning with the same spin magnitude 
and the spins initially parallel to each other ($\kappa_1=\kappa_2$), the parameter bias increases slightly to 
$\Delta \chi = 0.03$. Note that in all of these cases the match has a sharp peak at its
maximum, but the match at the theoretical $\chi_{\rm eff}$ value is well above 
$0.97$ in many cases. 

The results do not change appreciably when the calculations are repeated with the 
Advanced LIGO noise curves. The matches improve slightly as the
mass is increased, but so does the bias in $\chi_{\rm eff}$. However, the bias never 
increases by more than $\Delta \chi = 0.01$. 
The results for the $20M_{\odot}$ bin
are displayed in the last two columns of Tab. \ref{tab:pn1} and \ref{tab:pn2}. We would like to 
emphasize again that QA waveforms are \emph{not} in a detector frame: the matches
using the detector noise curves are only to rule out the possibility of spurious results with 
the white-noise curve. 

The second set was chosen such that all configurations have the same theoretical 
$\chi_{\rm eff}$-value, but that the amount of precession changes due to a varying 
$\kappa_1=\kappa_2 \equiv \kappa$ angle. 
All configurations in this set are equal-spinning, i.e., the spins are initially 
equal in magnitude and parallel to each other. The results are given in 
Tab.~\ref{tab:pn2}. 
We see for both mass ratios $q=3$ and $q=10$ that the best-match $\chi$ agrees
with $\chi_{\rm eff}$ for small $\kappa$. A bias appears as $\kappa$ increases 
beyond $30^\circ$, but is again never more than $\Delta \chi = 0.02$. 

It is important to note that the parameter that describes the rate of inspiral, i.e.,
the phasing of the binary, is given by Eq.~\ref{eqn:chieff} and that the geometric
quantity that defines the amount of precession is quantitatively described by
the spin components perpendicular to $\vec{L}$, which are proportional to
$\sin \kappa_i$. We have looked at various other cases with 
varying relative azimuth angle between the spin vectors as well as varying
relative inclination between $\vec{S}_1$ and $\vec{S}_2$, i.e. $\kappa_1\neq \kappa_2$.
For equal spin magnitudes we find that the azimuth has no effect on the best-match $\chi_{\rm eff}$.
For unequal $\kappa_i$ but equal spin magnitude we find that the best-match bias increases
with increasing $\kappa_i$ but that the relative inclination angle between the two
spin vectors does not have a significant influence on the results. 

The approximation that $\chi_{\rm eff}$ is constant becomes less accurate as the binary
approaches merger. 
Remarkably, the effective spin value
associated with the initial $\chi_{\rm eff}$ value seems to characterize the best-match
non-precessing-binary system in all cases. Even when using detector
noise curves and choosing masses such that the late inspiral (when $\chi_{\rm eff}$ 
changes fastest) is in the most sensitive part of the detector band, the best-match
$\chi_{\rm eff}$ varies by only $\Delta \chi \leq 0.04$ from the value predicted by our 
hypothesis. However, it is likely that when we move to full IMR configurations, 
some other appropriate effective total spin will be more appropriate, as was found 
for the full IMR waveforms in Ref.~\cite{Ajith:2009bn}.

When interpreting these results, one should bear in mind that the phasing of a PN
waveform can change significantly with respect to the choice of PN approximant. The
matches that we calculated between QA and non-precessing waveforms are in 
general far better than those between, for example, the same non-precessing configuration
produced with TaylorT1 and TaylorT4; see Fig.~6 in Ref.~\cite{Ajith:2012tt}. 
In this sense, our approximation can be considered
to hold, well within the level of accuracy of our PN waveforms. 

\begin{table*}[t]
 \centering 
\begin{tabular}{c|c|c|c|c|c|c}
\hline
   q & $\vec{\chi}_1$  & $\vec{\chi}_2$   & $\chi_{\rm BM}$ &  $\mathscr{M}_0$ & $(\chi_{\rm BM})_{early}$ & $(\chi_{\rm BM})_{zdethp}$ \\
   \hline
   3 & $(0,0,0)$ & $(0.75,0,0)$  & $0.02$ & $0.9815$ & $0.02$ & $0.02$ \\
   3 & $(0.75,0,0)$ & $(0,0,0)$  & $0.00$ & $0.9997$ & $0.00$ & $0.00$ \\
   3 & $(0.75,0,0)$ & $(0.75,0,0)$ & $0.03$ &$0.9576$ & $0.04$ & $0.03$ \\
   \hline
   10 & $(0,0,0)$ & $(0.75,0,0)$ &  $0.03$ & $0.8209$ & $0.03$ & $0.03$  \\
   10 & $(0.75,0,0)$ & $(0,0,0)$ & $0.00$ & $0.9999$ & $0.00$ & $0.00$ \\
   10 & $(0.75,0,0)$ & $(0.75,0,0)$ & $0.03$ &$0.8075$ & $0.03$ & $0.03$ \\
    \hline
\end{tabular}
  \caption{PN configurations with constant $\kappa_i=90^\circ$ for the spinning hole and varying spins. The best matches, not necessarily for the predicted $\chi_{\rm eff}=0$ but for the values displayed in column 4, are all well above 0.999 for $q=3$ and above $0.995$ for $q=10$. $\mathscr{M}_0$ denotes the match with the counterpart waveform that has $\chi_{\rm eff}=0$. The last two columns show the best match for two potential Advanced LIGO noise curves, evaluated for a $20M_\odot$ binary. For all cases the best match is above $0.999$ for both detector noise curves.} 
  \label{tab:pn1} 
\end{table*}

\begin{table*}[top]
 \centering 
\begin{tabular}{c|c|c|c|c|c|c}
\hline
   q & $\vec{\chi}_1=\vec{\chi}_2$  & $\kappa_1=\kappa_2$ & $\chi_{\rm BM}$  &  $\mathscr{M}_{0.5}$  &$(\chi_{\rm BM})_{early}$ & $(\chi_{\rm BM})_{zdethp}$   \\
   \hline
   3 & $(0.050,0,0.572)$  & $ 5^\circ$  & $0.50$ & 0.9998 & $0.50$ & $0.50$  \\
   3 & $(0.101,0,0.572)$  & $10^\circ$ & $0.50$ & 0.9998 & $0.50$ & $0.50$  \\
   3 & $(0.208,0,0.572)$  & $20^\circ$ & $0.50$ & 0.9992 & $0.51$ & $0.50$ \\
   3 & $(0.330,0,0.572)$  & $30^\circ$ & $0.51$  & $0.9975$ & $0.51$ & $0.51$ \\
   3 & $(0.480,0,0.572)$  & $40^\circ$ & $0.52$  & $0.9917$ & $0.52$ & $0.52$ \\
   3 & $(0.682,0,0.572)$  & $50^\circ$ & $0.52$  & $0.9719$ & $0.52$ & $0.52$  \\
   \hline
   10 & $(0.093,0,0.529)$  & $10^\circ$  & 0.50 &  0.9986 & $0.50$ & $0.50$  \\
   10 & $(0.193,0,0.529)$  & $20^\circ$  & 0.50 &  0.9996 & $0.50$ & $0.50$  \\
   10 & $(0.306,0,0.529)$  & $30^\circ$  & 0.50 &  0.9965 & $0.51$ & $0.50$  \\
   10 & $(0.444,0,0.529)$  & $40^\circ$  & 0.51 & $0.9771$ & $0.51$ & $0.51$  \\
   10 & $(0.631,0,0.529)$  & $50^\circ$  & 0.52 & $0.8925$ & $0.53$ & $0.52$ \\
   \hline
\end{tabular}
  \caption{PN configurations with the same effective spin value $\chi_{\rm eff}=0.5$ but varying $\kappa_1=\kappa_2$ for the two mass ratios $1:3$ and $1:10$. $\chi_{\rm BM}$ denotes the effective $\chi_{\rm eff}$-value yielding the best match. 
  In all cases the best matches are above 0.999 for $q=3$ and above 0.997 for $q=10$. $\mathscr{M}_{0.5}$ denotes the match with the counterpart waveform that has $\chi_{\rm eff}=0.5$. Column 5 lists the match for the predicted $\chi_{\rm eff}$-value. The last two columns show the best match for two potential Advanced LIGO noise curves, evaluated for a $20M_\odot$ binary.}
  \label{tab:pn2} 
\end{table*}

We also emphasize once again that the QA waveforms do not correspond to the waveforms as seen by 
a detector, since the QA frame is accelerating, and would not be
directly employed in a GW search; the matches as shown therefore do not constitute 
a study of the efficacy of these waveforms for either searches or parameter estimation.
What they do tell us, however, is that if we were to take the non-precessing waveforms
used in this study, and to apply the reverse QA procedure to them, i.e., ``wrap them
up'' into mock precessing waveforms using the inverse QA angles calculated for
each of these configurations, then we would expect them to agree well with the original
precessing-binary waveforms. This study also suggests that if we were to construct a
waveform model from ``wrapped up'' non-precessing waveforms, then it is 
possible that this model could
be used to measure the effective total spin $\chi_{\rm eff}$ with only a small 
bias. However, the true behavior of such a model in a parameter estimation exercise
requires an exhaustive study that is beyond the scope of this paper. 

\begin{figure}
\begin{center}
\includegraphics[width=80mm]{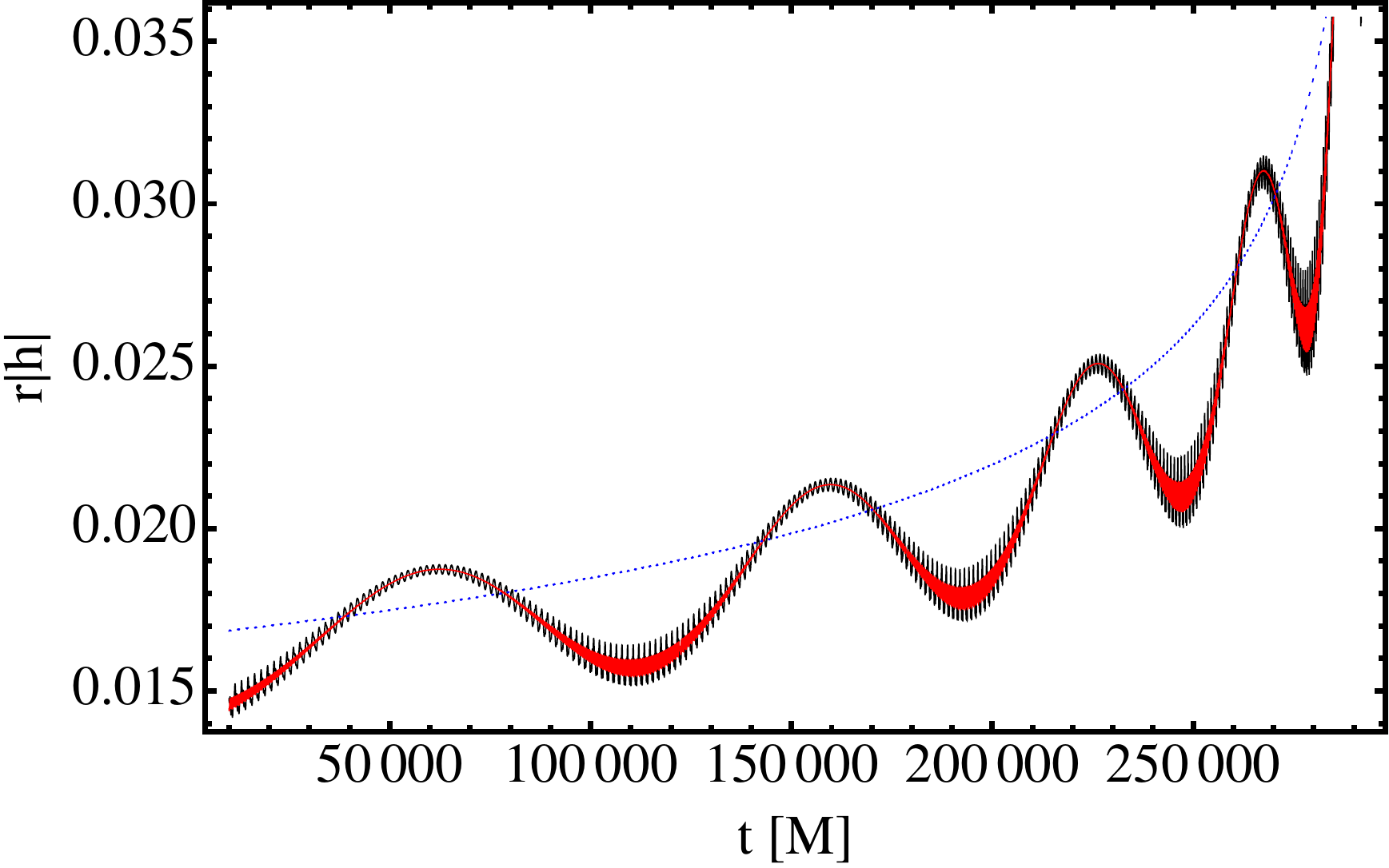}
\caption{
The absolute value of the GW strain for a precessing binary, as viewed 
at an arbitrary inclination of 2.8\,rad from $\hat{J}_0$. The signal includes all
$\ell =2$ modes. The true signal (black) has the finer structure; the other signal
with the lower-amplitude high-frequency oscillations (red) was generated by twisting a non-spinning $q=3$ waveform
with the inverse QA angles. The dotted line shows the 
amplitude of the original non-spinning waveform.
}
\label{fig:mockprecession}
\end{center}
\end{figure}

To back up this claim, we performed the following exercise: from the first case in 
Tab.~\ref{tab:pn1} we took the $\chi_{\rm eff}$ waveform, which is non-spinning 
$q=3$, and wrapped it up with the reverse QA angles that we calculated for the 
$\{q=3, \chi_1 = 0, \chi_{2,x} = 0.75\}$ configuration. 
The resulting waveform is 
shown in Fig.~\ref{fig:mockprecession}; we have plotted the absolute value of the
GW strain, constructed from all $\ell=2$ modes, at an arbitrarily chosen 
inclination of 2.8\,rad from the initial direction of the total angular momentum,
$\hat{J}_0$. Also
shown is the same quantity for the ``true'' precessing-binary waveform, and for
comparison we also show the original non-precessing-binary waveform, constructed
from only the $(\ell=2,|m|=2)$-modes. We see
that the twisted-up non-precessing-binary waveform captures the main features
of the amplitude of the true precessing-binary waveform extremely well; how well
the phases agree can be judged by calculating the match between the two waveforms. 
This we did, once again over the frequency range of $fM \in [0.0018,0.01]$.
Note that now we \emph{are} considering waveforms as they would be observed
in a detector.

We find that the match between the true precessing-binary waveform
and the mock-precession waveform have a match greater than $0.97$ for all 
masses and binary orientations. By contrast, the match between the 
unmodified non-precessing $q=3$ waveform and the true precessing waveform is below 0.97 even for the 
best-performing orientation.  
These results provide an important cross-check that we can indeed 
mimic the original PN precessing-binary signal by suitably transforming the
signal from a non-precessing binary. 

As an aside, note that there is one mode of the precessing-binary signal that
we cannot fully model in this way, the $(\ell=2,m=0)$-mode. In the non-precessing
waveforms, the $(2,2)$- and $(2,-2)$-modes are complex conjugates of each other. 
When this is true, the transformed $(2,0)$-mode will always be real. This can
be seen from inspection of Eq.~(\ref{eq:h22QA}). But in the 
true precessing-binary waveform the $(2,0)$-mode has real and imaginary parts;
it is straightforward to produce an example to illustrate this from Sec.~IV of
Ref.~\cite{Arun:2008kb}. In order to capture these effects, we would need to
break the symmetry between the non-precessing $h_{\ell m}$ modes, which 
would require that our model include unequal spins --- this is therefore one
limitation of a single-effective-spin model. In practice, however, the relative signal
power in the \emph{imaginary part} of the $(2,0)$-mode (that part that our model cannot
reproduce) will always be small, 
and we expect the other errors in this approximate waveform, for example in the phasing, 
will be more significant 
in practice.

\subsection{Transitional Precession}
\label{sec:transitional}
In the previous section we have seen that our mapping works extremely well in cases of 
simple precession; in fact it can be considered to be an exact mapping within the 
error bars of the PN phasing. In this section, we demonstrate that it also works in the 
more extreme case 
of transitional precession~\cite{Apostolatos:1994mx}. This second type of precession occurs 
when $\vec{L}$ and $\vec{S}$ are almost opposite and equal in magnitude and so 
$|\vec{J}|$ is small. During the 
inspiral, the magnitude of $\vec{S}$ hardly changes but since orbital angular momentum 
is radiated away, the magnitude of $\vec{L}$ decreases with time. With the appropriate 
choice of parameters, the total angular momentum $\vec{J}$ is initially small and positive, 
but due to the loss of orbital angular momentum, decreases until it crosses the $xy$-plane 
of the Cartesian source frame, where it changes sign. 
See Ref.~\cite{Apostolatos:1994mx} for an extensive discussion of transitional precession.

\begin{figure*}
\begin{center}
\vspace{-1cm}
\includegraphics[width=70mm]{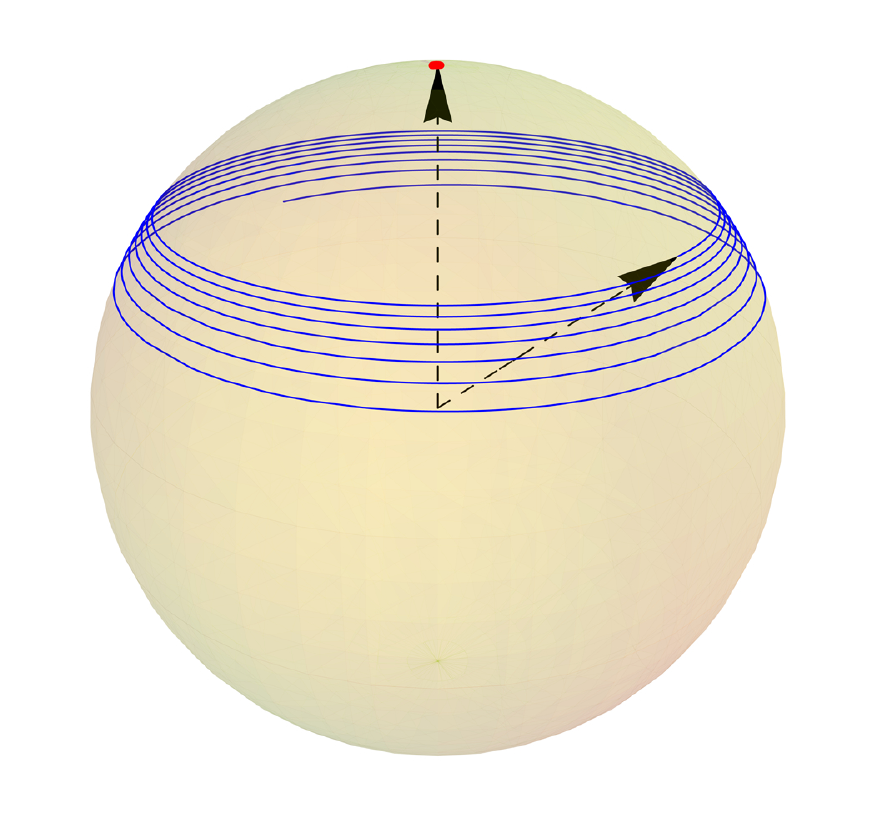}
\includegraphics[width=70mm]{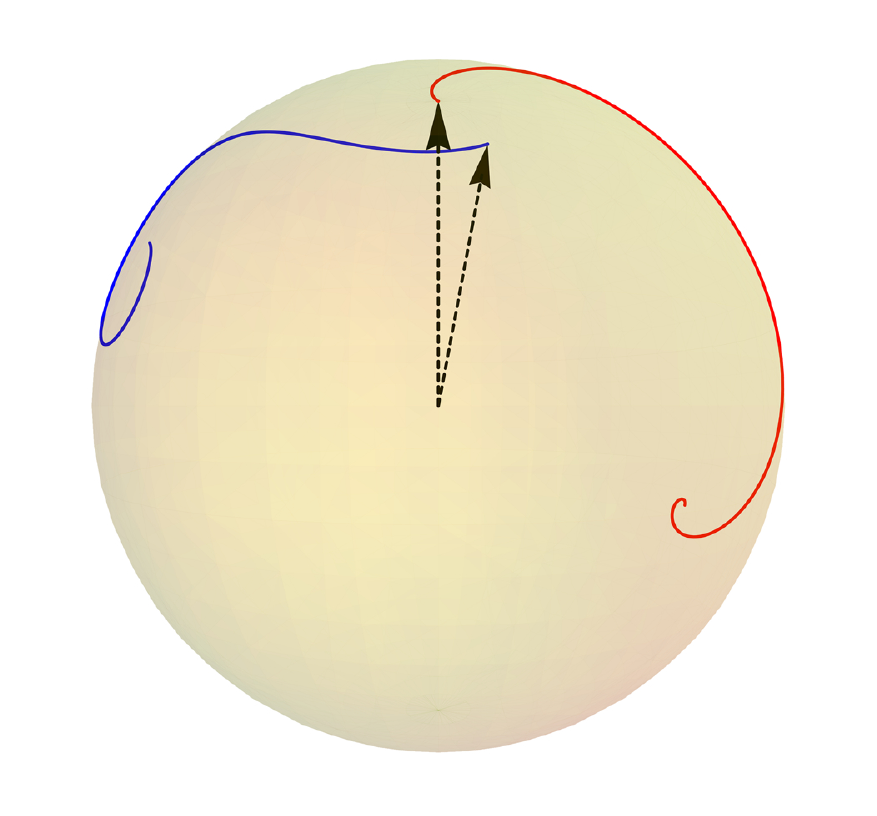}
\caption{Evolution of $\vec{J}$ (red) and $\vec{L}$ (blue) plotted on the unit sphere, where $\vec{J}$ 
is initially aligned with the direction $(0,0,1)$. The left panel shows the evolution of these two 
directions for a case of simple precession. The precession cone described by $\hat{J}$ is very 
small in comparison to the one described by $\hat{L}$, and appears on the figure as only a dot at the end of the vertical arrow. The right panel shows the same characteristic
directions for a case of transitional precession. In this case $\vec{J}$ clearly moves along the unit 
sphere away from its initial direction (to the right side of the sphere) and separates from $\vec{L}$, which moves to the left side of the sphere in the figure.}
\label{fig:JLmotion}
\end{center}
\end{figure*}

As opposed to simple precession, where $\hat{J}_0$ represents the least evolving axis 
in the binary's geometry, this direction changes significantly during the transitional phase, 
as shown in Fig.~\ref{fig:JLmotion}. In order to test the validity of our 
precessing$\rightarrow$non-precessing mapping for a transitional-precession 
case, we have chosen one specific 
configuration with PN parameters $q=10$, initial separation $D_i=53$M, and initial spins 
$\vec{\chi}_1=(0,0,0)$ and $\vec{\chi}_2=0.65\cdot(0,-\sin(3^\circ),-\cos(3^\circ))$. 
This is a single-spin configuration, where the initial spin is $3^\circ$ from complete 
anti-alignment and the generated inspiral waveform is about $2 \cdot10^6M$ long, 
terminating at a final separation of $D_f=6M$. 

It is worth mentioning that in order to produce a transitional phase, the parameters have 
to be fine-tuned such that $\vec{J}$ changes sign. If $\vec{S}$ and $\vec{L}$ were 
completely anti-aligned, no precession would occur at all. The transitional phase
is not brief: it takes up most of the duration of the inspiral that we have calculated,
and, as noted in Ref.~\cite{Apostolatos:1994mx}, cases where a binary undergoes
transitional precession within the sensitivity band of ground-based detectors 
are expected to be very rare.  

\begin{figure}
\begin{center}
\includegraphics[width=80mm]{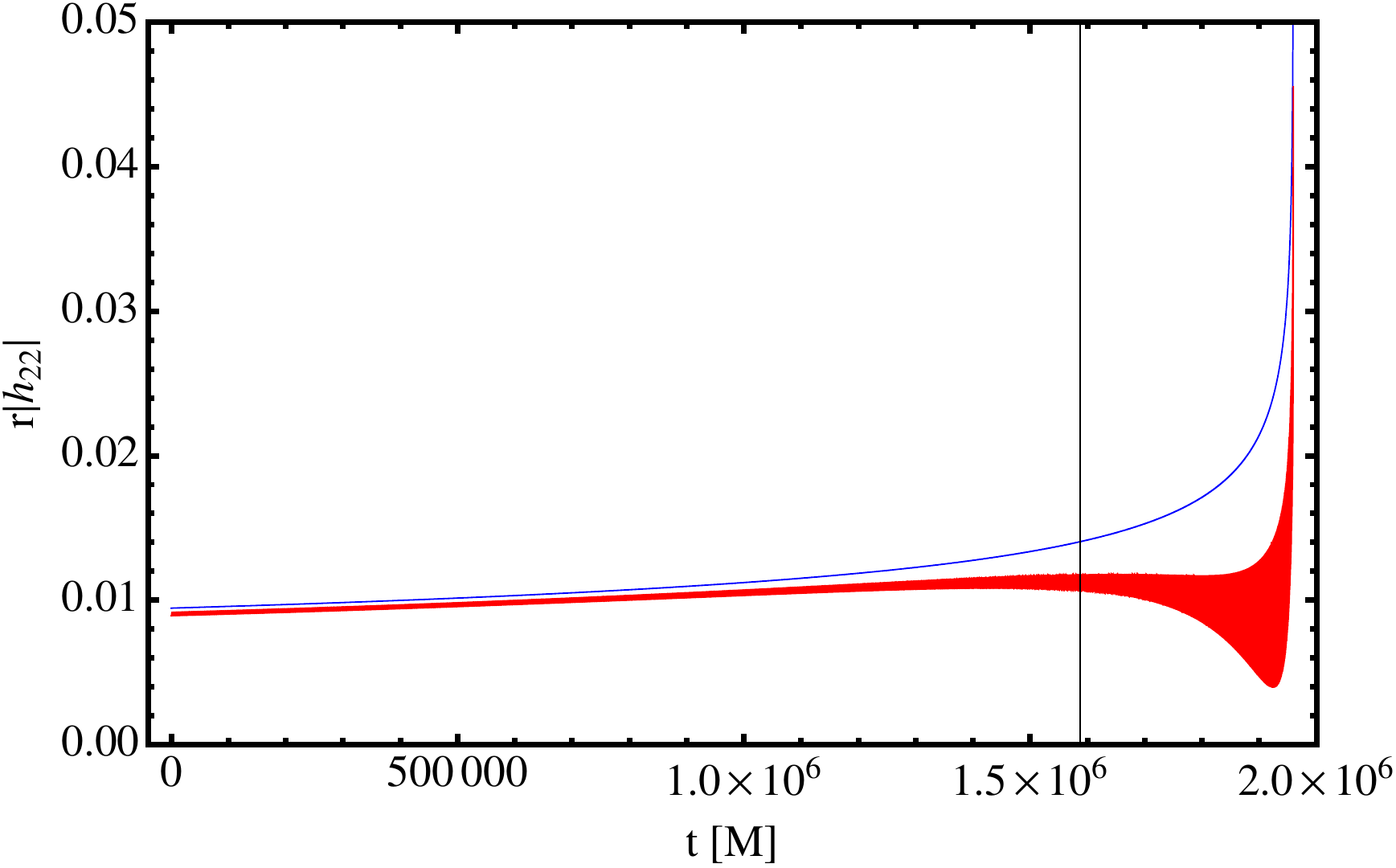}
\caption{The panel shows the magnitudes of the $(2,2)$ modes for the transitional 
precession case before (red; lower curve) and after (blue; upper curve) the quadrupole alignment was applied. 
The change of the direction of $\hat{J}$ at $t=1.587\cdot 10^6M$ is indicated by the 
vertical line. A strong modulation is introduced into the original waveform at that time, 
which is completely removed after quadrupole alignment.}
\label{fig:h22transition}
\end{center}
\end{figure}

The dramatic change of the direction of $\vec{J}$ is reflected in the GW signal and 
the transitional waveforms in the standard source frame look particularly distorted 
when the total angular momentum crosses the $xy$-plane, as is shown in Fig.~\ref{fig:h22transition}. 

\begin{figure}[t]
\begin{center}
\includegraphics[width=80mm]{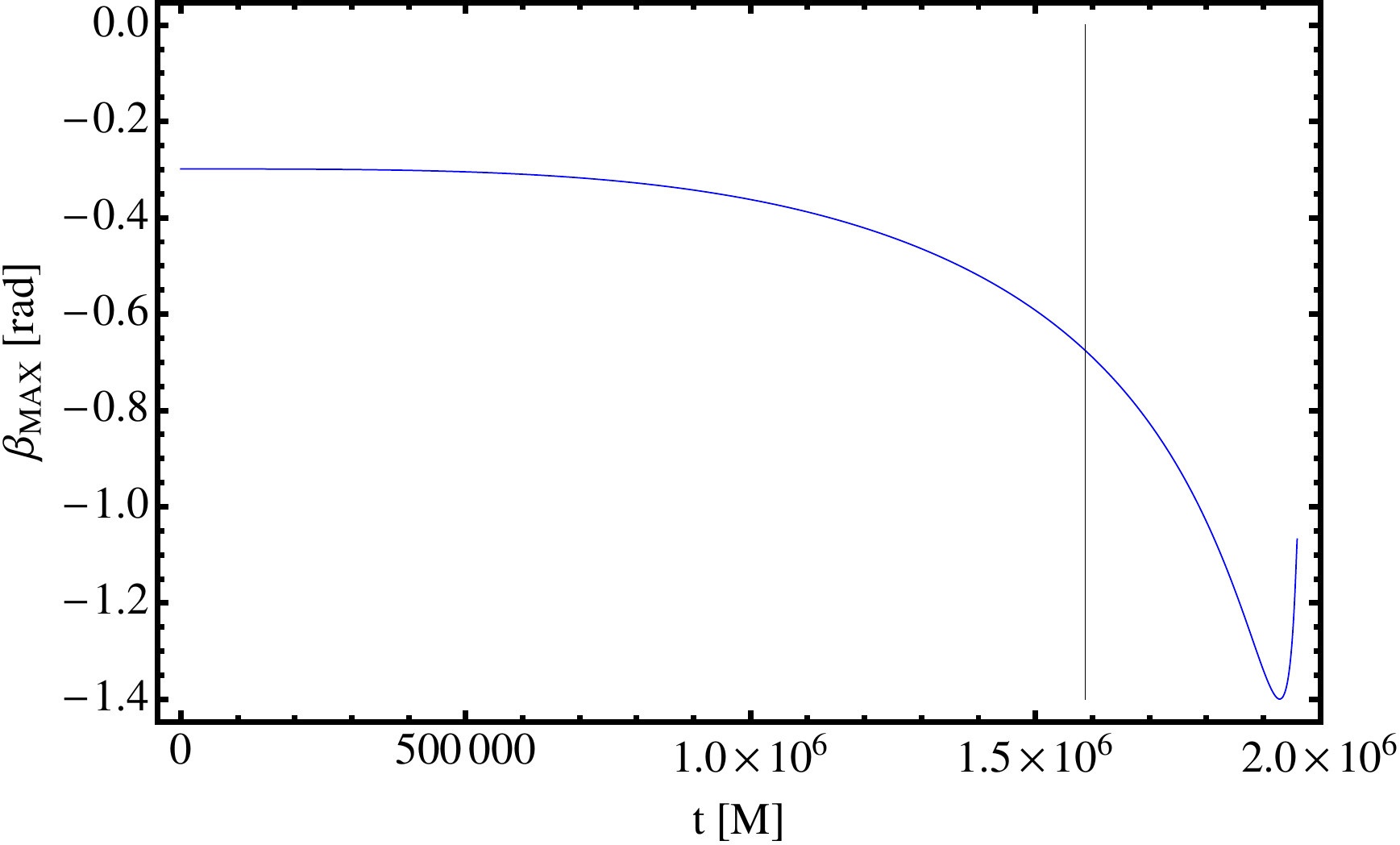}
\includegraphics[width=80mm]{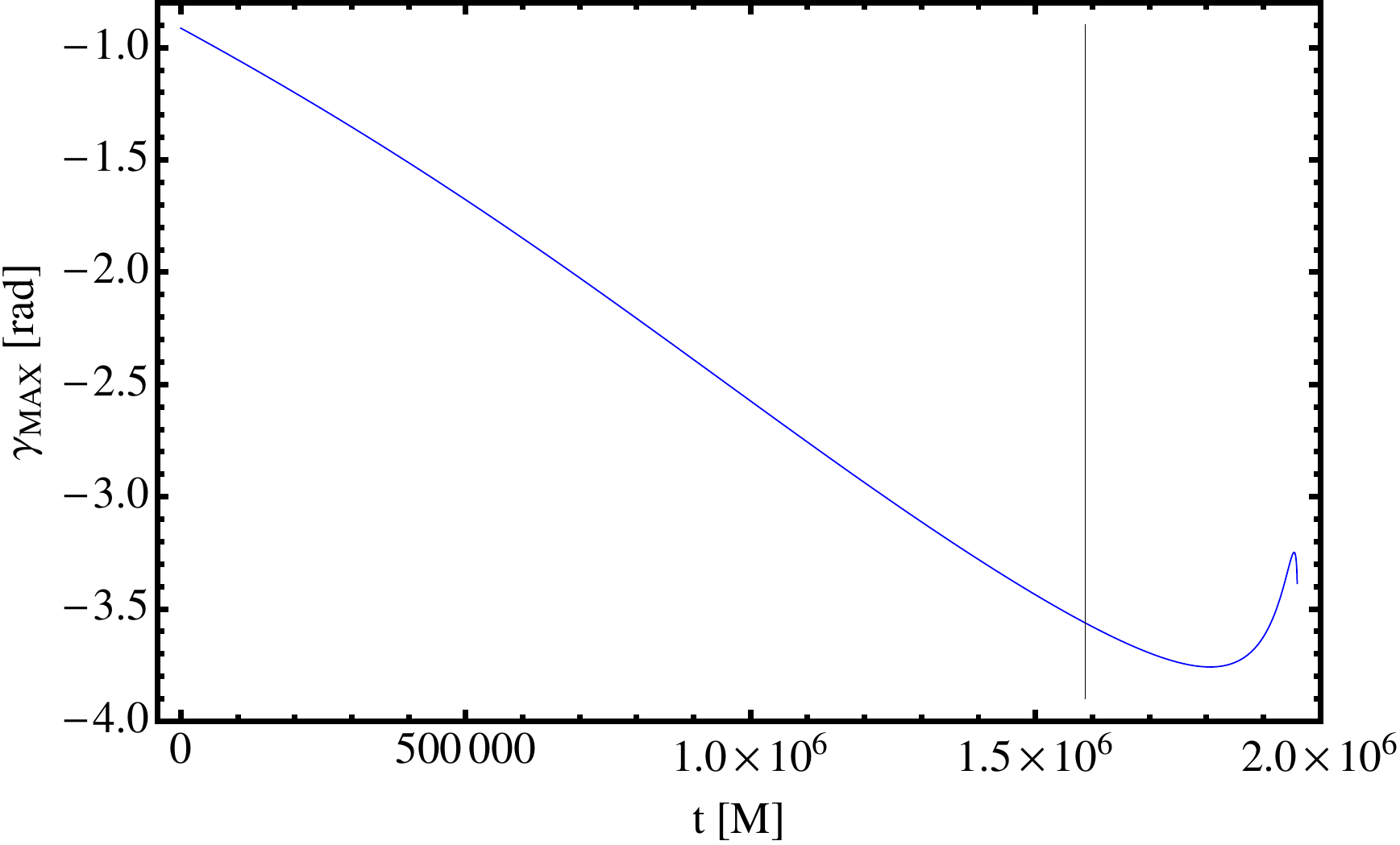}
\caption{The two panels show the two Euler angles $\beta$ and $\gamma$ determined 
by the quadrupole-alignment procedure for the transitional case. The time when the 
$z$-component of $\vec{J}$ changes sign is indicated by the vertical line.}
\label{fig:transAngles}
\end{center}
\end{figure}

We do not expect any of these features 
to be present in the quadrupole-aligned waveform, since we now track the direction of 
dominant emission, and this is completely independent from any asymptotic direction of $\vec{J}$.
We see in Fig.~\ref{fig:h22transition} that this is indeed the case. 
The angles found by the maximization 
routine are shown in Fig.~\ref{fig:transAngles}. The zero-crossing of
the total angular momentum occurs at $t=1.587\cdot10^6M$, which is indicated in the figures with a
vertical line. 

If our hypothesis were correct, then the QA waveform would be very close to a non-precessing 
waveform with $\chi_{\rm eff} =-0.572$, from Eq.~(\ref{eqn:chieff}). As before, we compared 
the QA mode with 
a series of non-precessing waveforms with varying spin parameter to locate the non-precessing 
configuration that agrees best with the QA waveform. We find the best match to be 
$0.998$ for a spin anti-aligned waveform with effective 
spin parameter
$\chi_{\rm eff}=-0.576$.
 This is remarkably close to the theoretically expected value, with a bias of only
 $\Delta \chi = 0.004$!  

On the other hand, naively using the non-aligned transitional-precession waveform and 
calculating the matches with the same comparison waveforms gives the same effective 
spin value, since the phase is dominated by the inspiral rate, but yields a best match of 
only $0.940$. Note also that this is for the $(2,2)$-mode as seen from only one orientation;
for many other orientations that matches are likely to be far worse. 

This example demonstrates that even in the case of transitional precession, our method 
proves to be accurate (expected $\chi_{\rm eff}$-value) and robust 
($\mathscr{M} > 0.99$) for mapping precessing waveforms onto single-spin-parameterized 
non-precessing-binary waveforms.

\section{PN-NR hybrid waveforms}
\label{sec:hybrids}

So far we have discussed only PN inspiral waveforms. To produce complete
waveforms that include the late inspiral, merger and ringdown, we need to 
include results from numerical-relativity (NR) simulations. In this section we
will show how the quadrupole-alignment procedure simplifies the production 
of hybrid PN-NR waveforms. 

A variety of methods have been introduced to construct hybrid waveforms for
non-precessing 
configurations~\cite{Pan:2007nw,Ajith:2007qp,Ajith:2007kx,Boyle:2009dg,Santamaria:2010yb,Hannam:2010ky},
and see Ref.~\cite{Ajith:2012tt} for a unified summary of the methods in use. 
In all methods the PN and NR waveforms
are aligned at some time, or over a time or frequency window, and then blended
together. Such waveforms have been used to 
produce phenomenological waveform 
models~\cite{Ajith:2007qp,Ajith:2007kx,Ajith:2007xh,Ajith:2009bn,Santamaria:2010yb}, 
and are now also being
used to test GW search and parameter estimation tools~\cite{Ajith:2012tt}. A 
number of studies have also been performed on the length requirements of
NR waveforms in order to produce sufficiently physically accurate 
hybrids~\cite{Hannam:2010ky,MacDonald:2011ne,Boyle:2011dy,Ohme:2011zm}
and these also include estimates of the influence of errors and 
ambiguities in the hybridization process on the physical fidelity of the final
waveform. 

The construction of hybrids for precessing-binary configurations is more 
complex: not only do the time and phase of the PN and NR waveforms 
have to be aligned, but to some extent the orientations of the spins and 
orbital plane must agree as well. For the precessing-binary hybrids 
that were used in Ref.~\cite{Ajith:2009bn}, the hybrid waveforms were constructed 
by matching the NR waveforms with PN waveforms computed from the same 
PN evolution that was employed to construct the initial data for the NR simulations.
This technique ignores mismatches in the binary orientation and physical 
parameters due to the emission of junk 
radiation~\cite{Hannam:2006zt,Lovelace:2008hd} 
and gauge changes~\cite{Hannam:2006vv,Hannam:2008sg} in the early 
stages of an NR simulation, although these effects are expected to be small;
see Ref.~\cite{Santamaria:2010yb} for a detailed discussion of this point in the
context of non-precessing-binary hybrids. 

These complications can be avoided through the use of QA
waveforms. The PN and NR waveforms, both converted to the QA frame, 
can now be aligned exactly as in the non-precessing cases. In order to 
reverse the QA process, it is also necessary to align the QA angles 
$(\beta, \gamma, \epsilon)$, but this is straightforward, as we show below. 

In the next section we will outline how we produce a QA hybrid for the 
precessing-binary waveform that we used in Ref.~\cite{Schmidt:2010it}.
This corresponds to the 
first configuration discussed in Tab.~\ref{tab:pn1}: 
$q=3$, $\chi_1 = 0$, $\chi_2 = 0.75$,
and $\vec{S}\cdot\vec{L} = 0$. 
Having produced the QA hybrid, we will
examine where our non-precessing-binary mapping hypothesis breaks
down as we approach merger. That the hypothesis must break down is clear,
because the spin of the final merged black hole will be influenced by the 
black-hole spins in a way that the orbital phase evolution is not, and the 
mass and spin of the final black hole will \emph{not} be the same as that 
for the corresponding non-precessing inspiral configuration. 
 
\subsection{Construction of QA hybrids}
\label{sec:makehybrid}

A QA hybrid can be produced by making use of the same procedure as for a non-precessing-binary
hybrid. We will briefly summarize the method that we used. 

We start with a PN and an NR waveform, each for the same physical configuration. The last
requirement is achieved to good approximation by using results from the PN evolution to 
produce the initial parameters for the NR evolution. The PN and NR waveforms are then 
put into the QA frame by the procedure described in Sec.~\ref{sec:qa}. We will produce a hybrid
of $\Psi_4$, and note that, since the QA frame is non-inertial, we cannot produce 
$\Psi_4^{QA}$ by taking two time derivatives of $h^{QA}$. We must first produce 
the $\Psi_{4,2m}$ modes from the original precessing-binary GW-strain modes, $h_{2m}$, and 
apply the QA algorithm to $\Psi_{4,2m}$.

\begin{figure}[h!]
\begin{center}
\includegraphics[width=80mm]{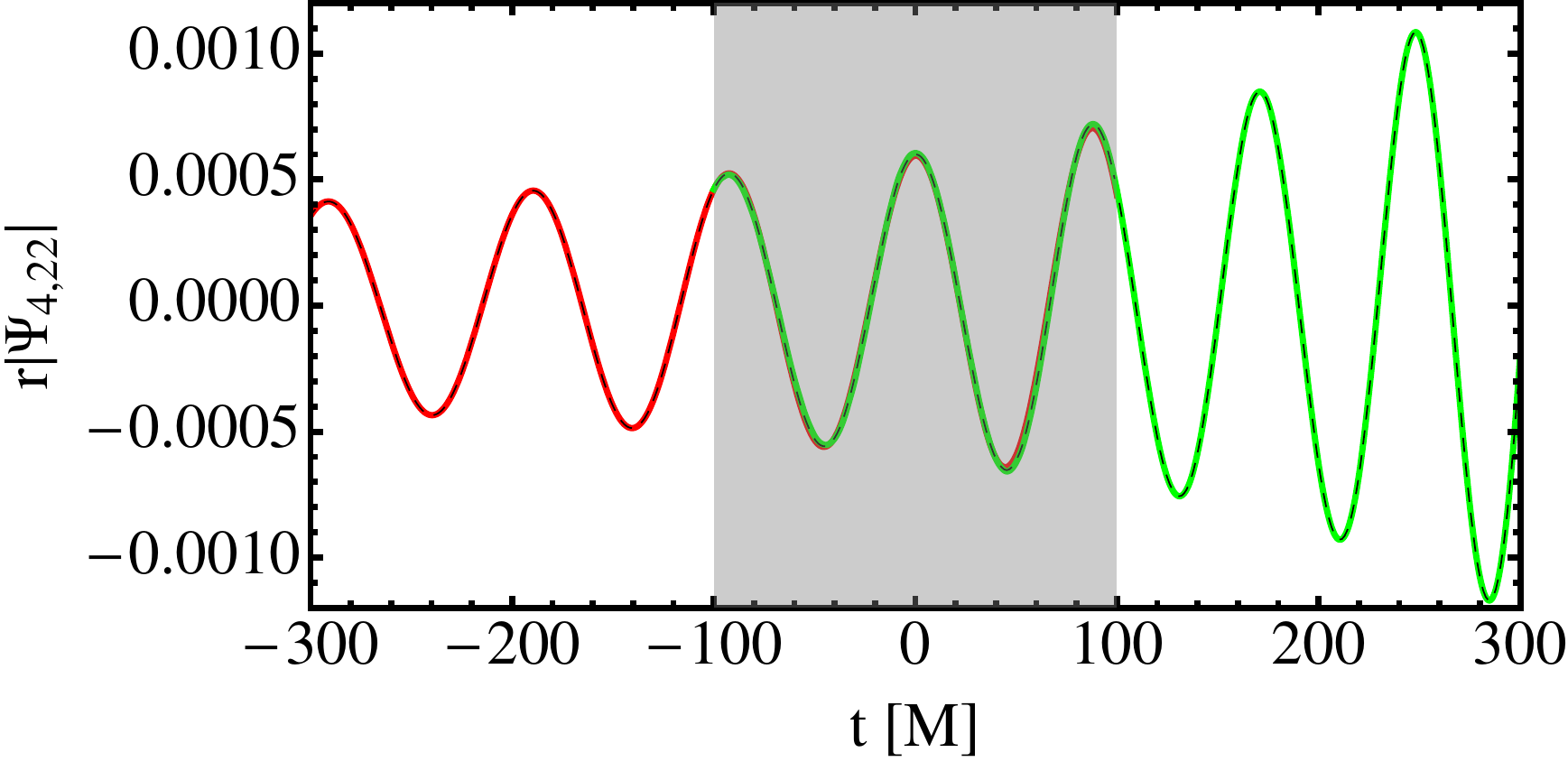}
\caption{
The PN (red, from $t=-300M$ to $t=100M$), NR (green, from $-100M$ to $300M$) and hybrid (dashed black) waveforms near the matching time 
($t=0$). The PN and NR waveforms are blended together in the window $\Delta t = [-100,100]$,
indicated by the shaded region.
}
\label{fig:hybrid}
\end{center}
\end{figure}

We then choose a matching
frequency $\omega_m$, and locate the times $t_{PN}$ and $t_{NR}$ when each 
waveform passes through that frequency. For our $q=3$ configuration, $\omega_m = 0.07$.
We align the PN and NR frequencies around that time such that both
$\phi_{PN}(t_{PN}) = \phi_{NR}(t_{NR})$ and $\omega_{PN}(t_{PN}) = \omega_{NR}(t_{NR})
= \omega_m$. The hybrid waveform is then produced by blending together 
$\Psi_{4,PN}^{QA}$ and $\Psi_{4,NR}^{QA}$ with a linear transition function of width
$\Delta t = 200\,M$ around the matching frequency. The final waveform is then 
\begin{equation}
\Psi_{4,hyb}^{QA}(t) = a_- \Psi_{4,PN}^{QA}(t - t_{PN}) + a_+ \Psi_{4,NR}^{QA}(t-t_{NR}),
\end{equation} where $a_{\pm} = (\Delta t/2 \pm t)/\Delta t$ when $t \in [-\Delta t,\Delta t]$ and
zero or one otherwise, and the time has been 
shifted such that $t=0$ coincides with the point at which $\omega = \omega_m$. This
constitutes the QA hybrid. Fig.~\ref{fig:hybrid} shows the real part of $\Psi_4$ around the
time where the matching was performed, which is at $t=0$. The figure shows the
PN and NR waveforms, as well as the final hybrid, and we see that the matching between
the PN and NR waveforms is smooth.

\begin{figure}[h!]
\begin{center}
\includegraphics[width=80mm]{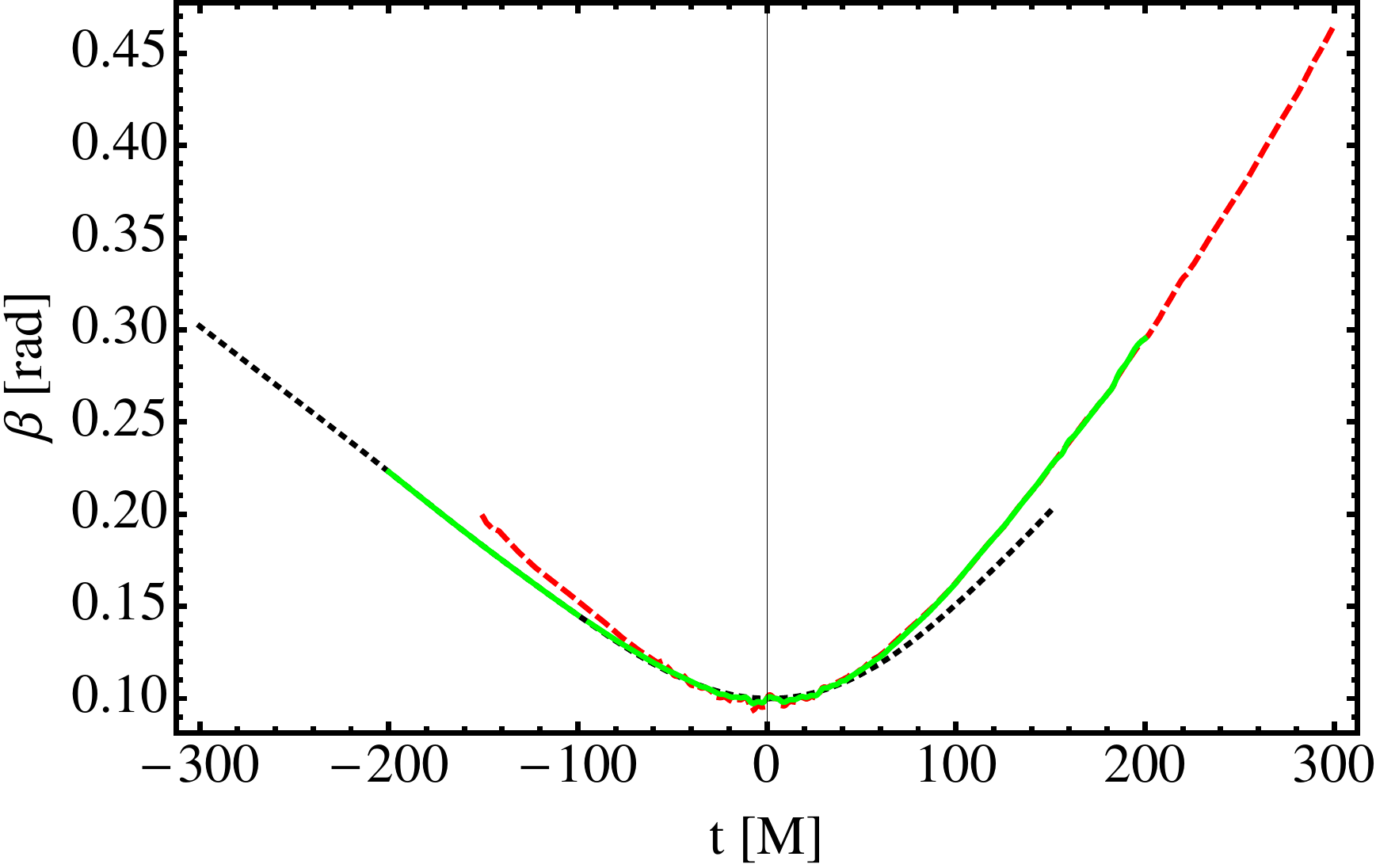}
\includegraphics[width=80mm]{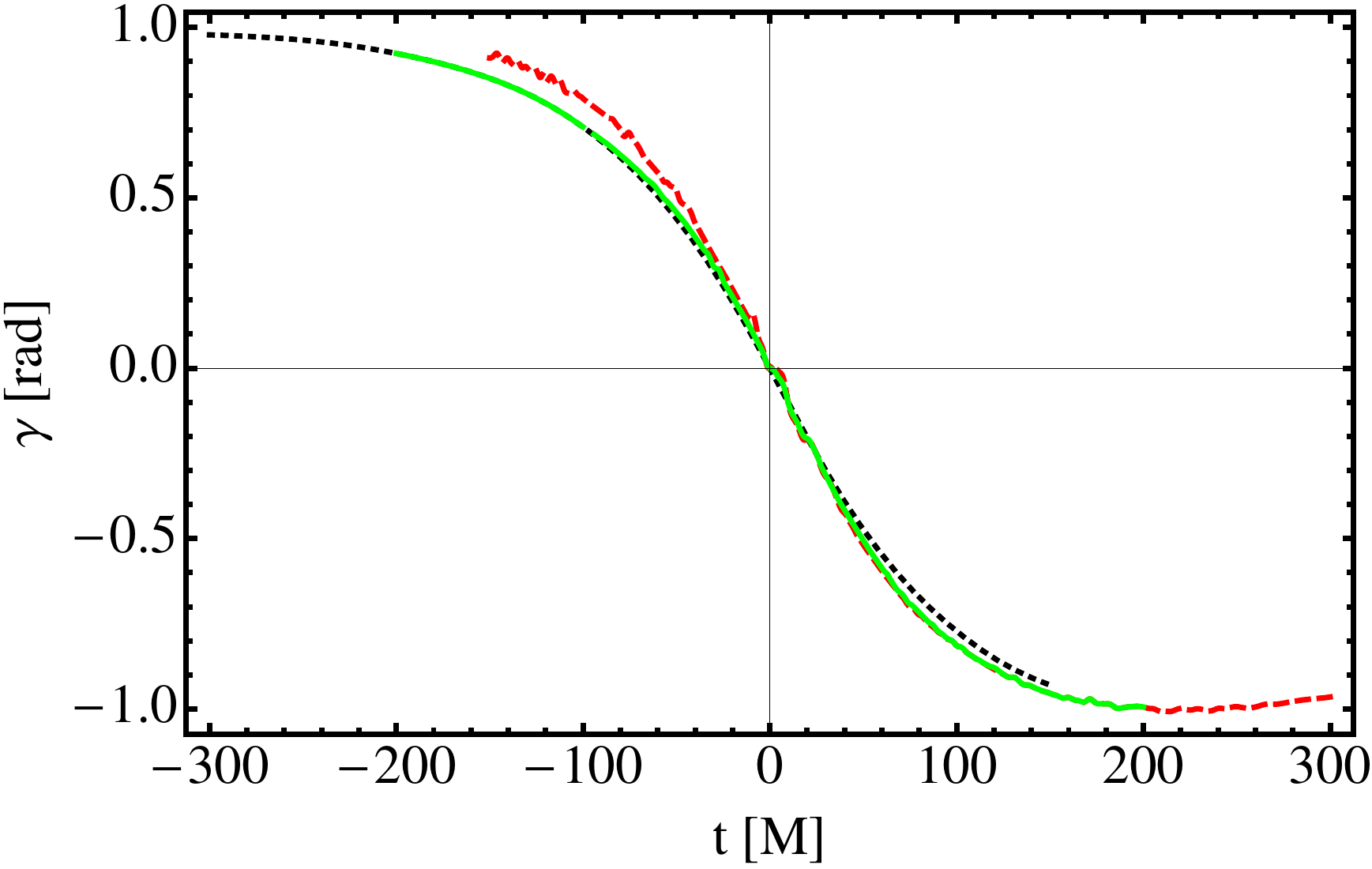}
\includegraphics[width=70mm]{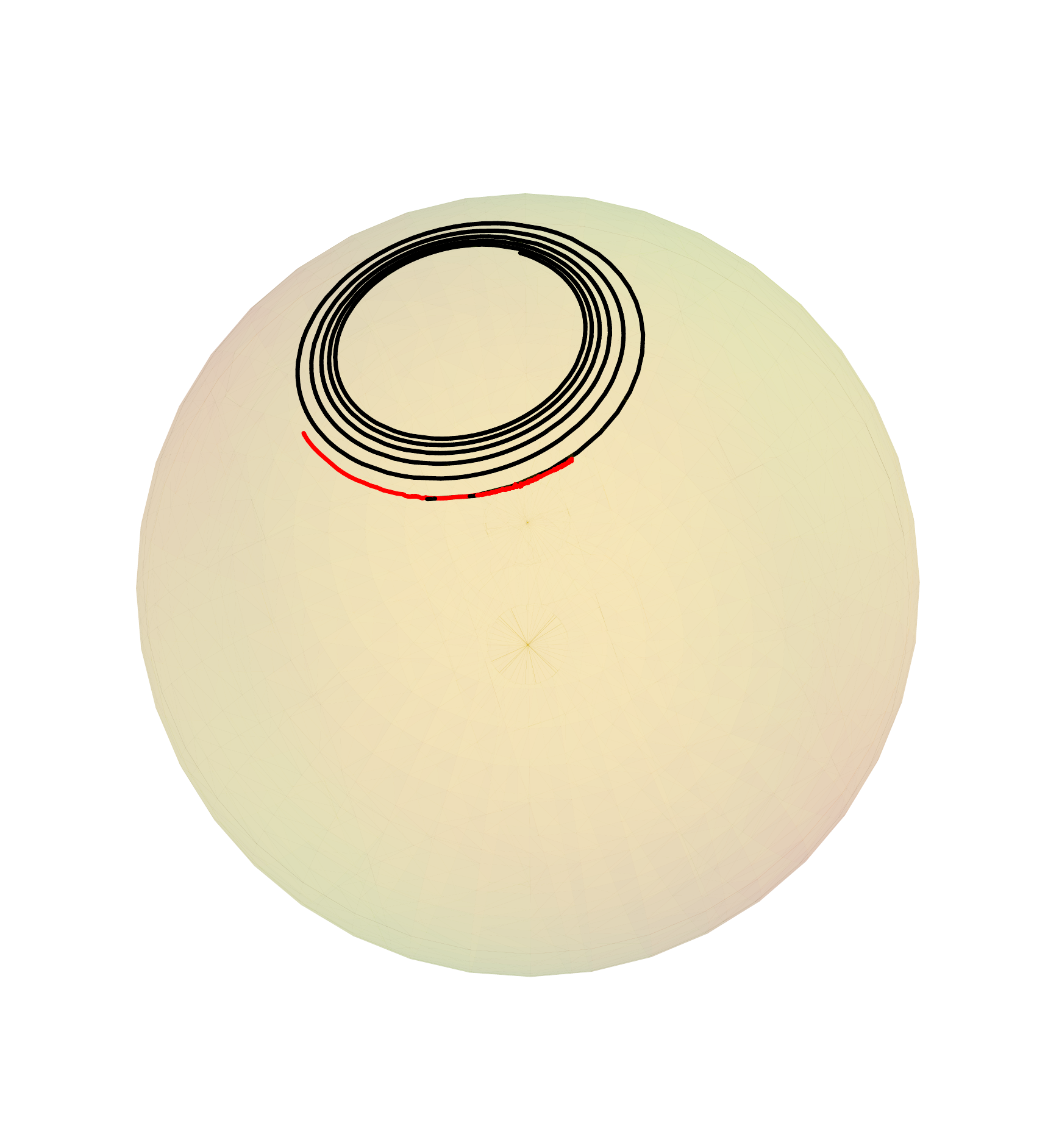}
\caption{Hybridization of the QA angles $\beta$ and $\gamma$. Upper two panels:
the black (dotted) lines indicate
the inspiral PN values, the red (dashed) lines indicate the later NR values, and the green (solid) lines
indicate the hybrids. The lower figure shows the evolution of the aligned QA directions, where here the 
black line indicates long PN inspiral of duration $2.9\times10^5\,M$, and the red line indicates 
the NR results up to merger.}
\label{fig:hybangles}
\end{center}
\end{figure}

To convert this into a physical precessing-binary hybrid, we also require hybrids 
of the QA angles 
$\{\beta(t), \gamma(t),\epsilon(t)\}$. These are produced as follows. The two angles 
$\{\beta(t), \gamma(t)\}$ define a vector $\vec{n}(t) = \{\sin(-\beta(t))\cos(-\gamma(t)), 
\sin(-\beta(t))\sin(-\gamma(t)),\cos(-\beta(t))\}$, which traces out a path on the unit 
sphere. The QA angles for the PN waveform define $\vec{n}_{PN}(t)$, while those for
the NR waveform define $\vec{n}_{NR}(t)$. We perform a fixed rotation $\mathbf{R}_{PN}$ to 
$\vec{n}_{PN}(t)$ (and another $\mathbf{R}_{NR}$ to $\vec{n}_{NR}(t)$), such that both vectors
are equal at the matching frequency, $\vec{n}_{PN}(t_{PN}) = \vec{n}_{NR}(t_{NR})$. Since
the angle $\gamma$ is ill-defined when $\vec{n} = \{0,0,1\}$, we do not choose that
as our (arbitrary) matching direction, but rather the vector such that 
$\beta(t_{PN}) = 0.1\,$rad. 
Specification of a third Euler angle allows us to require that
the vectors not only meet at the matching time, but that the curves they trace out
are parallel at that time. To do this we simply measure the angle between the two 
curves at the matching time, and then rotate $\vec{n}_{NR}(t)$ around 
the axis defined by the matching direction, $\vec{n}_{NR}(t_{NR})$.
Fig.~\ref{fig:hybangles} shows the first two angles at the times close to 
the matching frequency, and the final PN and NR curves are shown in the 
lower panel of Fig.~\ref{fig:hybangles}.
 The hybrid angles are constructed by smoothly 
blending between the PN and NR angles, in the same way as for the 
QA waveform. 
The precessing-binary hybrid can then be constructing by performing the reverse
QA procedure with $\{\gamma, \beta, \epsilon\} \rightarrow \{-\epsilon, -\beta, -\gamma\}$.

\subsection{Breakdown of the non-precessing-binary equivalence}
\label{sec:hybridcomp}

We expect the simple mapping between QA- and non-precessing-binary waveforms to break
down near merger. As we have seen, the effect of the spins on the inspiral rate comes 
predominantly from the spin components parallel to the orbital angular momentum; this is
why our mapping works. At merger, however, the spin of the final black hole is, to first 
approximation, $\vec{J}_{fin} = \vec{L} + \vec{S}_1 + \vec{S}_2$, where the orbital and spin
angular momentum vectors are those at the point of merger. (A far more sophisticated treatment
of the final spin ingredients is given in Ref.~\cite{Buonanno:2007sv}, and a number of
estimates of the final spin as a function of the initial configuration exist in the 
literature~\cite{Rezzolla:2007rd,Tichy:2008du,Lousto:2009mf}.)
All components of the 
spin now become important and the appropriate parameterization may no longer be the 
effective total spin $\chi_{\rm eff}$.

\begin{figure}
\begin{center}
\includegraphics[width=80mm]{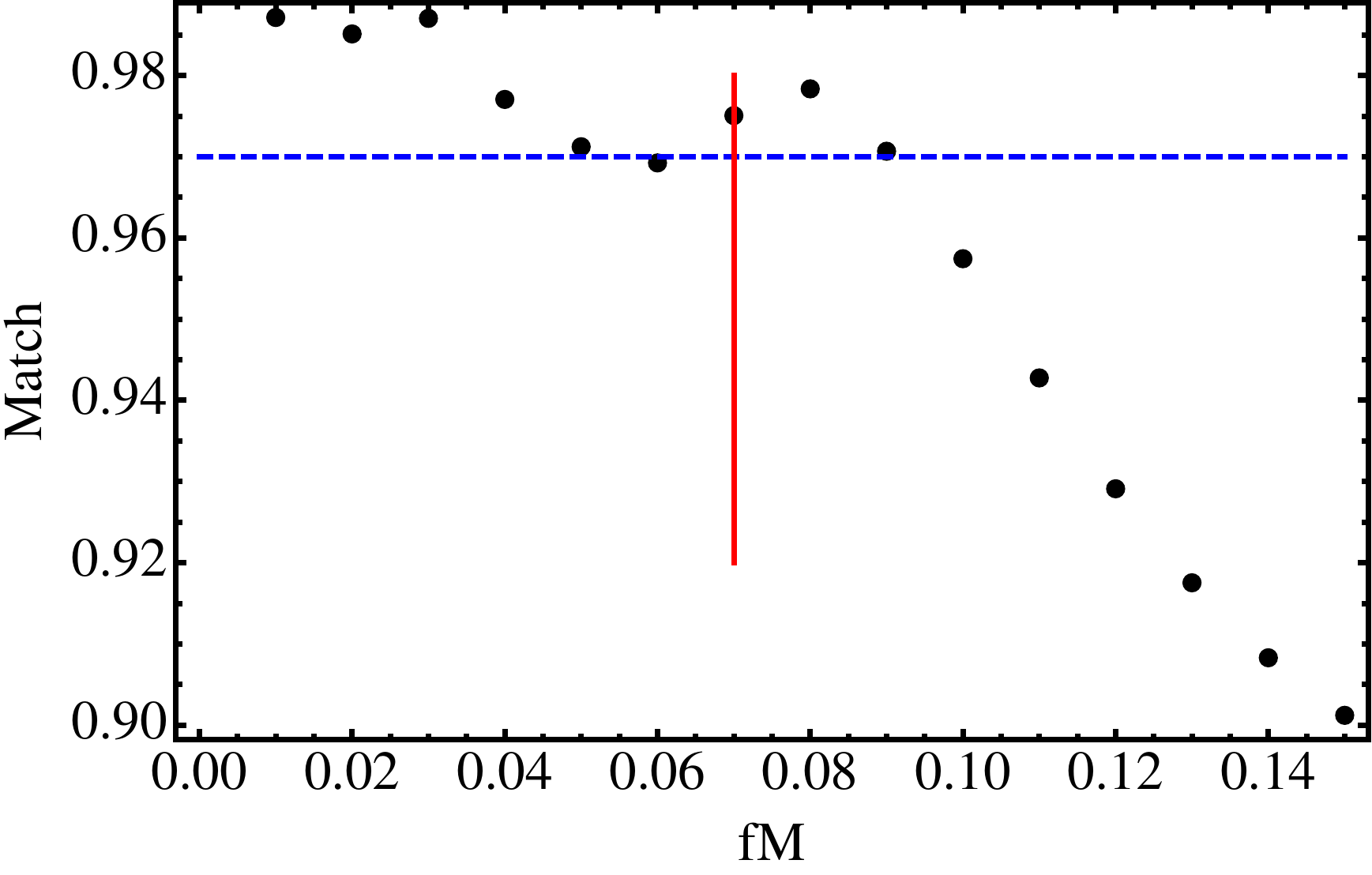}
\caption{Matches between QA and non-precessing hybrids, for our standard $q=3$ configuration.
The horizontal axis represents the frequency at which both waveforms are cut off in the match calculation,
and indicates that the two hybrids agree well (match $> 0.97$) right up to the merger, indicated
by the vertical line.}
\label{fig:hybmatch}
\end{center}
\end{figure}

It is instructive to investigate where the mapping breaks down, and we can use the hybrid
waveform constructed in the previous section to do this. Fig.~\ref{fig:hybmatch} shows the 
match between the QA hybrid constructed above, and a non-spinning $q=3$ hybrid (which 
would be the appropriate non-precessing configuration during the inspiral). The match is
calculated for a range of termination frequencies of the two waveforms. For reference, 
the frequency $fM = 0.016$ corresponds roughly to $M\omega = 0.1$, and is close to the point
where PN waveforms are typically terminated in inspiral searches. Below this frequency the 
white-noise match is consistent with the results in Sec.~\ref{sec:simple}. 
The peak of the waveform occurs
at $fM = 0.07$, which is indicated by a vertical line. The fiducial acceptable match of $0.97$ is
indicated by a horizontal line. We see that the match is at or above $0.97$ through the merger,
and only degrades significantly during the ringdown. 

Once again we emphasize that these matches were computed using a white-noise power spectrum.
Nonetheless, these provide evidence that the QA procedure is valid very close to the merger, and perhaps
even up to ringdown. We will discuss the implications of this result for waveform modeling in 
the final section. 

\section{Discussion: A route to generic-binary waveform models}
\label{sec:discussion}

We have extended previous work on the quadrupole-alignment (QA) procedure to show that 
it can be used not only to cast precessing-binary waveforms in a simple form, but to map 
these waveforms onto a sub-family of non-precessing waveforms. We verified that this 
sub-family is parametrized by only mass ratio and an effective total spin parameter, and that the
non-precessing waveform that best matches each QA waveform (with white-noise 
matches of at least $0.995$), corresponds to our predicted $\chi_{\rm eff}$ value to within 
$\Delta \chi \leq 0.04$. The mapping was tested on a range of inspiral PN waveforms with 
mass ratios $q=3$ and $q=10$, and even on an example of transitional precession; in all cases
the approximations holds well within the level of accuracy of the PN phasing. 
As a final test, we used the reverse QA procedure to ``wrap up'' a non-precessing-binary waveform,
and found that it matched the corresponding true precessing waveform with a match of 
$>0.97$ for all binary orientations. 
We also 
showed that this procedure can simplify the construction of hybrid PN-NR waveforms, and
that the approximate mapping seems to hold all the way through to merger. 

Our results suggest that generic precessing-binary waveforms can be generated with good accuracy 
by applying the reverse of the quadrupole-alignment transformation to a small class of 
non-precessing-binary waveforms. These waveforms appear to faithfully represent the ``true'' 
precessing-binary waveforms up to the point of merger, and perhaps even up to 
the ringdown. The
problem of constructing a generic waveform model can then be factorized into the smaller problem of
modeling the two rotation angles $\{\beta(t),\gamma(t)\}$ as a function of the black-hole spins
and the mass ratio. 

More concretely, we propose the following strategy: once the evolution of the Euler angles 
$\beta(t)$ and $\gamma(t)$ has been determined for a large sample of the configuration space, 
these can be modeled as functions that depend on some set of parameters 
$\vec{\lambda}$
\begin{eqnarray}
\beta & = & \beta(\vec{\lambda}(t)), \\
\gamma & = & \gamma(\vec{\lambda}(t)). 
\end{eqnarray} 
We emphasize that the $\vec{\lambda}$ should be \emph{physical} parameters, or a combination of 
physical parameters. 
The third angle $\epsilon(t)$ is automatically determined given the two others. The rotation 
angles are unique up to an overall rotation of the frame of reference; we expect that they will 
assume the simplest form if in the limit of infinite binary 
separation $\hat{J}_{-\infty} = (0,0,1)$.

Since precessing inspiral-merger (IM) waveforms can be mapped onto 
non-precessing ones via quadrupole alignment, using the angles 
$\{ \gamma(t), \beta(t), \epsilon(t)\}$, a phenomenological IM 
model with (anti-)aligned spins can be used and ``twisted up'' with the inverse angles
$\{ -\epsilon(t), -\beta(t), -\gamma(t)\}$. 
This will give us a phenomenological IM model, 
\begin{equation}
\label{ }
h_{lm}^{IM}(t) = \mathbf{R}(-\epsilon,-\beta,-\gamma) \, h_{lm}^{IM}(\eta, \chi_{\rm eff}; t).
\end{equation} 

Needless to say, an \emph{inspiral} model is not urgently needed: we can already produce generic
waveforms by integrating the PN equations of motion, as we have done in this paper. 
A simple closed-form approximation to these solutions could significantly improve the efficiency of 
gravitational-wave search and parameter-estimation pipelines, but there is no barrier in principle to 
producing theoretical estimates of any of these signals. The real need is for complete IMR 
waveform models. 

Given in addition a phenomenological model for the ringdown, 
$h_{\ell m}^R(\vec{\lambda}_R;t)$, which is parameterized by some yet-to-be determined
subset $\vec{\lambda}_R$ of the full binary parameters $\vec{\lambda}$, we expect that
we can produce
a combined IMR model, which can be indicated schematically as
\begin{equation}
\label{ }
h_{lm}^{IMR}(t) = \mathbf{R}(-\epsilon,-\beta,-\gamma) \, h_{lm}^{IM}(\eta, \chi_{\rm eff}; t)
\times h_{\ell m}^R (\vec{\lambda}_R; t).
\end{equation} 
For ease of use in GW searches, ideally this model would be cast in
closed-form expressions in the frequency domain. 

We still have the problem of modeling a seven-dimensional parameter space, but we now have
to model only two functions, and, as we can see from Fig.~\ref{fig:QAangles} 
(and even Fig.~\ref{fig:transAngles} for transitional 
precession), they are smooth, simple
functions, that may be far easier to model than the complicated amplitude and phase modulations
that are standard features of the physical waveforms. It is also likely that many of the features of 
the full seven-dimensional parameter space can be captured by a model that considers only a subset
of the parameters. It is also quite possible that we will need to employ a non-precessing model 
that treats both black-hole spins, and/or the effective spin that proves most useful will differ from that
presented here. Our main purpose is only to outline a general research program to develop 
generic waveform models based on QA waveforms.

In the short term, a number of issues deserve further study. One is that the QA method, in all forms 
proposed to date~\cite{Schmidt:2010it,OShaughnessy:2011fx,Boyle:2011gg}, 
has only been studied in detail prior to merger. 
It is likely that in order to apply the QA method optimally through the ringdown, it will be advantageous
to make use of spheroidal (rather than spherical)
harmonics, but that remains to be seen. Another will be the appropriate blending between the 
inspiral-merger regime and the ringdown, which is likely to be parameterized by the final mass and spin, 
and probably also the mass ratio and a second effective spin parameter~\cite{Kamaretsos:2012bs}.
It would also be instructive to explore the effectiveness of such signals for both GW searches and 
parameter estimation across a wide volume of the binary parameter space. 

We consider all of these to constitute a promising strategy for the construction of approximate 
generic waveform models, and will pursue them further in future work.


\section*{Acknowledgments}

We thank G. Faye for providing us with a Mathematica notebook containing all 
waveform-mode expressions used in this work.
We also thank P. Ajith, S. Fairhurst, I. Harry, F. Ohme, M. P\"urrer and B. Sathyaprakash for discussions; 
and A. Buonanno, E. Ochsner and F. Ohme for comments on the manuscript.
P. Schmidt is a recipient of a DOC-fFORTE-fellowship of the Austrian Academy of Sciences
and was also partially supported by the STFC.
M. Hannam was supported Science and Technology Facilities Council grants ST/H008438/1
and ST/I001085/1.
S. Husa was supported by
grant FPA-2007-60220 from the Spanish Ministry of Science and
the Spanish MICINN’s Consolider-Ingenio 2010 Programme under grant 
MultiDark CSD2009-00064, and thanks Cardiff University for hospitality.

{\tt BAM} simulations were carried out at Advanced Research Computing 
(ARCCA) at Cardiff, LRZ Munich, the Vienna Scientific Cluster (VSC), 
MareNostrum at Barcelona Supercomputing Center -- Centro Nacional de 
Supercomputaci\'on (Spanish National Supercomputing Center), and on the 
PRACE clusters Hermit and Curie.


\bibliography{ModellingPaper.bib}

\end{document}